\documentclass[fleqn,usenatbib]{mnras}

\usepackage[T1]{fontenc}

\DeclareRobustCommand{\VAN}[3]{#2}
\let\VANthebibliography\thebibliography
\def\thebibliography{\DeclareRobustCommand{\VAN}[3]{##3}\VANthebibliography}

\def\equationautorefname#1#2\null{(#2)}

\usepackage{graphicx}	 
\usepackage{amsmath}	 
\usepackage{amssymb}	 
\usepackage{mathtools}   
\usepackage{CJK}         
\usepackage{xcolor}      
\usepackage{scalerel}
\usepackage{tikz}
\usetikzlibrary{svg.path}
\definecolor{orcidlogocol}{HTML}{A6CE39}
\tikzset{
  orcidlogo/.pic={
    \fill[orcidlogocol] svg{M256,128c0,70.7-57.3,128-128,128C57.3,256,0,198.7,0,128C0,57.3,57.3,0,128,0C198.7,0,256,57.3,256,128z};
    \fill[white] svg{M86.3,186.2H70.9V79.1h15.4v48.4V186.2z}
                 svg{M108.9,79.1h41.6c39.6,0,57,28.3,57,53.6c0,27.5-21.5,53.6-56.8,53.6h-41.8V79.1z M124.3,172.4h24.5c34.9,0,42.9-26.5,42.9-39.7c0-21.5-13.7-39.7-43.7-39.7h-23.7V172.4z}
                 svg{M88.7,56.8c0,5.5-4.5,10.1-10.1,10.1c-5.6,0-10.1-4.6-10.1-10.1c0-5.6,4.5-10.1,10.1-10.1C84.2,46.7,88.7,51.3,88.7,56.8z};
  }
}

\usepackage{newtxtext,newtxmath}

\newcommand{\lz}[1]{#1}

\newcommand\orcidicon[1]{\href{https://orcid.org/#1}{\mbox{\scalerel*{
\begin{tikzpicture}[yscale=-1,transform shape]
\pic{orcidlogo};
\end{tikzpicture}s
}{|}}}}

\title[Split-Monopole Accretion Column]{Dynamics of Neutron Star Accretion Columns in Split-Monopole Magnetic Fields}

\author[L. Zhang et al.]{
Lizhong Zhang (张力中) \orcidicon{0000-0003-0232-0879},$^{1}$\thanks{E-mail: lizhong@physics.ucsb.edu}
Omer Blaes \orcidicon{0000-0002-8082-4573},$^{1}$
Yan-Fei Jiang (姜燕飞) \orcidicon{0000-0002-2624-3399}$^{2}$
\\
$^{1}$Department of Physics, University of California, Santa Barbara, CA 93106, USA\\
$^{2}$Center for Computational Astrophysics, Flatiron Institute, New York, NY 10010, USA\\
}

\date{Accepted 2023 January 1. Received 2022 December 20; in original form 2022 October 11}

\pubyear{2022}

\begin{document}
\begin{CJK*}{UTF8}{gbsn}
\label{firstpage}
\pagerange{\pageref{firstpage}--\pageref{lastpage}}
\maketitle

\begin{abstract}
We perform 2D axisymmetric radiative relativistic MHD simulations of radiation pressure supported neutron star accretion columns in split-monopole magnetic fields.  The accretion columns exhibit quasi-periodic oscillations, which manifest in the luminosity power spectrum as $2-10$~kHz peaks, together with broader extensions to somewhat higher frequencies.  The peak frequency decreases for wider columns or higher mass accretion rates.  In contrast to the case of shorter columns in uniform magnetic fields, \textit{pdV} work contributes substantially to maintaining the radiation pressure inside the column against sideways radiative cooling.  This is in part due to the compression associated with accretion along the converging magnetic field lines towards the stellar surface.  Propagating entropy waves which are associated with the slow-diffusion photon bubble instability form in all our simulations.  Radial advection of radiation from the oscillation itself as well as the entropy waves is also important in maintaining radiation pressure inside the column.  The time-averaged profile of our fiducial simulation accretion is approximately consistent with the classical 1D stationary model provided one incorporates the correct column shape.  We also quantify the porosity in all our accretion column simulations so that this may also in principle be used to improve 1D models.
\end{abstract}



\section{Introduction}
\label{sec:introduction}

An accretion-powered X-ray pulsar is a binary system with a strongly magnetized, rotating neutron star as its accretor.  The accretion disc in these systems is truncated near the Alfv\'en radius by strong magnetic fields anchored in the neutron star (see e.g. \citealt{Walter2015} for recent review).  In the vicinity of this truncation region, the magnetic pressure becomes dominant and constrains the accretion flow to fall along the magnetic field lines.  When the accretion rate is low, the infalling gas is shocked near the stellar surface and forms a hot spot, radiating in what has come to be called a pencil-beam pattern \citep{Basko1975, Basko1976}.  When the accretion rate is sufficiently high, a shock forms above the stellar surface. A radiation pressure dominated columnar structure is then formed below the shock front and radiates mostly from its sides in a fan-beam pattern \citep{Inoue1975, Basko1976}.  In either case, the misalignment of the radiation emitting polar regions with respect to the stellar rotation axis produces the observed pulsations (see e.g. \citealt{Caballero2012} for recent review).  Similar columnar structures appear to exist in pulsating ultraluminous X-ray sources (ULXs; see e.g. \citealt{Kaaret2017} for a review).  These make up a sizable fraction ($\sim 25\%$) of all ULXs \citep{RodriguezCastillo2020}.

An early analytical solution of the physical structure of neutron star accretion columns was proposed by \citet{Basko1976}, where the model was assumed to be 1D and stationary with a simplified top-hat column geometry.  Later studies refined this 1D stationary model by using more careful treatments of radiative transfer and magnetic opacities (e.g. \citealt{Mushtukov2015,West2017,West2017b}). However, since the magnetically confined accretion column is supported by radiation pressure against gravity, it is generally thought to be inherently unstable to the growth of entropy waves, a phenomenon which is also known as the `photon bubble instability' \citep{Arons1992,Gammie1998,Begelman2001,Blaes2003,Begelman2006}, even though it has nothing to do with bubbles or buoyancy.  All numerical simulations of accretion columns suggest that they are not stationary but instead extremely dynamical \citep{Klein1989, Klein1996, Kawashima2020, Paper2}.

This is the third in a series of papers exploring the local dynamics of neutron star accretion columns through numerical simulations.  Following \citet{Hsu1997}, we first conducted numerical simulations of static, non-accreting atmospheres on a strongly magnetized neutron star in \citet{Paper1}. \footnote{Nonlinear simulations of photon bubble instability of a static atmosphere in the rapid diffusion regime have been performed by \citet{Turner2005}, but this is less relevant to neutron star accretion columns, because in the interior of the column the viscous length scale is larger than the scale at which the rapid diffusion regime is reached.} We demonstrated that the modes of linear slow-diffusion photon bubble instability grow fastest for wavelengths near the radiation viscous length scale ($\sim 10^3$~cm), which would require grid cell sizes of $\sim30$~cm to resolve.  Therefore, to fully resolve the photon bubble dynamics is computational expensive and nearly impossible for a global simulation of the accretion column with current resources.

We then turned to exploring photon bubbles in columns that are actually accreting.  We began by simulating columns at modest accretion rates \citep{Paper2}, so that they were short enough that a uniform magnetic field was an adequate approximation.  This then enabled us to use a Cartesian grid with uniform grid resolution everywhere.  All of the simulated accretion columns exhibited high-frequency oscillations and the \textit{pdV} work on the sinking material was negligible.  The oscillatory behavior originated from an inability of a 2D stationary column to resupply energy through radiation advection to locally balance the sideways cooling of the column, an aspect of the physics that 1D models have not captured.  In a resolution study, we demonstrated that the existence of the photon bubbles does not alter the fundamental oscillation dynamics in the accretion column, but only introduces additional spatial complexity to the column structure.

In this paper, we conduct more global simulations of accretion columns in split-monopole fields at higher accretion rates.\footnote{We choose a split-monopole field rather than a more strongly converging dipole or quadrupole field because it enables us to adopt field lines that are nearly along the spherical polar grid.  This helps avoid numerical issues associated with inversion of conservative to primitive variables in strong magnetic fields.  A split-monopole field still captures the essential geometric effect of an inward converging field, more and more strongly confining the accretion flow in the lateral direction.}  We jointly solve the relativistic MHD and frequency-integrated radiation transfer equations to explore the dependence of the column dynamics on various global parameters.  We find significant differences with our previous Cartesian simulations due both to the larger column height and the converging magnetic field lines.  The paper is organized as follows.  In \autoref{sec:numerical_method}, we summarize the numerical technique and simulation setup.  In \autoref{sec:results}, we describe our simulation results, including the build-up of the column, its oscillatory behavior, the emergent light curves, comparison with the 1D stationary model, the energetics of the column, and the characterization and behavior of entropy waves.  In \autoref{sec:discussion}, we discuss several numerical caveats that might influence the simulation outcomes, how entropy waves affect the dynamics, compare our work to previous work in the literature, and discuss the observational significance of our simulations. In \autoref{sec:conclusions}, we summarize our conclusions.

\section{Numerical Method}
\label{sec:numerical_method}

\subsection{Equations}
\label{sec:equations}

Neutron star accretion columns are horizontally confined against radiation pressure forces by strong magnetic fields.  These fields lead to Newtonian Alfv\'en speeds that exceed the speed of light in the low density, free-fall region.  A Newtonian MHD simulation would therefore be forced to use very small time steps to achieve numerical accuracy and stability.  We avoid this problem by using the radiative relativistic MHD techniques that we developed in our previous papers \citep{Paper1, Paper2}, but now adapted to spherical polar coordinates \citep{Stone2020, Jiang2021}.  The conservation laws are summarized below in the sequence of mass, momentum, and energy conservation in the gas, and radiative transfer: 
\begin{subequations}
\begin{align}
    &\partial_0(\rho u^0) + \nabla_j(\rho u^j) = S_{\mathrm{gr}1}
    \quad, 
	\label{eq:particle_conserv}
	\\
	&\begin{multlined}[t]
	\partial_0(w u^0u^i - b^0b^i)
	\\
	\mkern25mu + \nabla_j\left(w u^iu^j + \left(P_g+P_m\right)\delta^{ij} - b^ib^j \right) = S_{\mathrm{gr}2}^i - S_{r2}^i
    \quad, 
	\end{multlined}
	\label{eq:mom_conserv}
	\\
	&\begin{multlined}[t]
	\partial_0\left[w u^0u^0 - \left(P_g+P_m\right) - b^0b^0\right]
	\\
	\mkern150mu + \nabla_j(w u^0u^j - b^0b^j) = S_{\mathrm{gr}3} - S_{r3}
    \quad, 
	\end{multlined}
	\label{eq:energy_conserv}
	\\
	&\partial_0I + n^j\nabla_j I = \mathcal{L}^{-1}(\bar{S}_r)
    \quad. 
	\label{eq:rad_transfer}
\end{align}
\end{subequations}
Here $\rho$ is the gas density in the fluid rest frame and $\delta^{ij}$ is the Kronecker delta. Using velocity in units of the speed of light ($c=1$), the gas four-velocity is defined as $(u^0, u^i)=\Gamma(1, v^i)$, where $v^i$ is the gas three-velocity and $\Gamma=(1-v_j v^j)^{-1/2}$ is the Lorentz factor. Given the three-vector of magnetic field $B^i$ and fluid-frame gas pressure $P_g$, we can then define the four-vector of magnetic field $b^{\mu}$, magnetic pressure $P_m$, gas enthalpy $w_g$, and the total enthalpy $w$ as follows: 
\begin{subequations}
\begin{align}
    &b^0 = u_jB^j,\quad b^i = \frac{1}{\Gamma}(B^i + b^0u^i)
    \quad, 
    \\
    &P_m = \frac{1}{2}b_{\nu}b^{\nu}
    \quad, 
    \\
    &w_g = \rho +\frac{\gamma}{\gamma-1}P_g
    \quad, 
    \\
    &w = w_g +b_{\nu}b^{\nu}
    \quad, 
    \label{eq:total_enthalpy}
\end{align}
\end{subequations}
where the adiabatic index of idea gas $\gamma=5/3$ is adopted in our computation. Note that we use Latin and Greek indices to denote the three-vector (e.g. $i=1,2,3$) and the four-vector (e.g. $\mu=0,1,2,3$) components, respectively.  The radiation intensity field $I$ is defined in the lab frame, as well as the unit vector of the photon propagation direction $n^i$.  Since $\nabla_i$ is expressed in spherical polar coordinates ($r, \theta, \phi$), extra geometric terms are introduced by the advection terms in both the relativistic MHD equations and radiative transfer equations, which can be found in \citet{Stone2020} and \citet{Davis2020}, respectively.
The quantities $S_{\mathrm{gr1}}$, $S_{\mathrm{gr2}}^i$, $S_{\mathrm{gr3}}$ are gravitational source terms derived by reducing the full general relativistic conservation laws to the weak field limit (for details see \citealt{Paper1}).
Similarly, $S_{r2}^i$ and $S_{r3}$ are radiation source terms for momentum and energy, respectively, computed from the lab-frame moments of the radiation intensity \citep{Jiang2021}.  Finally, $\bar{S}_r$ consists of the fluid-frame emissivity, absorption and scattering source terms \citep{Paper1,Jiang2021}, with
$\mathcal{L}^{-1}$ being the Lorentz boost operator from the comoving to lab frame.  Here, we denote the radiation quantities in the comoving frame with overbars. 

For simplicity, our study of the column dynamics in the split-monopole field configuration assumes that electron scattering is Thomson, i.e. isotropic with no dependence on magnetic field and polarization.  Our computation of radiative transfer is frequency-integrated, assuming a blackbody spectrum and neglecting non-zero photon chemical potential effects.  In our next paper, we will further investigate the dynamical effects caused by magnetic opacity.  This paper aims instead to obtain insight on the dynamical effects introduced by geometric dilution (e.g. $r^2$ for the spherical polar geometry used here) and high accretion rates.

\subsection{Simulation Setup}
\label{sec:simulation_setup}

As shown in \citet{Paper1}, the slow-diffusion photon bubble instability is highly resolution-dependent.  We therefore first studied accretion column dynamics using a Cartesian grid where numerical resolution was spatially constant \citep{Paper2}.  However, we found there that the global characteristics of the accretion column (e.g. shock height and oscillation frequency) are almost independent of resolution and therefore photon bubbles do not alter the more global dynamics of the column. This at least partially alleviates any concern related to numerical convergence of the column dynamics introduced by changes in spatial resolution in curvilinear coordinate grids.  For this reason, we adopt an axisymmetric, 2D spherical polar geometry in this paper, with a radial (split-monopole) magnetic field in the initial condition.  This enables us to explore accretion column dynamics at higher accretion rates, but we still limit the column height to be comparable to or less than the stellar radius in order to be qualitatively valid compared with a more realistic dipolar geometry.  The accretion column in the simulations in this paper are axisymmetric mounds at the magnetic pole.  Unlike the 2D Cartesian column simulations in \citet{Paper2}, where the simulated column was a slice through a long, thin wall of accreting material, here with this axisymmetric 2D structure of the accretion column we compute the entire escaping radiation field.

\subsubsection{Global Parameters and Simulation Domain}
\label{sec:global_parameters_and_simulation_domain}

The grid cells are uniformly sampled in the $\hat{\theta}$-direction with a constant polar angle interval $\Delta\theta$. We maintain the same aspect ratio for all cells, so the radial width $\Delta r$ increases by a fixed factor $(1+\Delta\theta)$ between radially adjacent cells.  Although the gas properties are defined in the 2D $r$-$\theta$ plane in axisymmetry, the radiation intensity field is configured in 3D angular grids with 64 angle directions.

We adopt a neutron star mass $M_{\star}=1.4M_{\sun}$ and radius $R_{\star}=10^6$~cm for our calculations.  The global geometry of the column is specified by the polar angle column width $\theta_{\mathrm{c}}$ and maximum radius $r_{\mathrm{c}}$, which needs to be set sufficiently large to cover both the free-fall and sinking zones.  The accretion rate of the column can be parameterized by the local Eddington ratio $\epsilon$, which is the ratio of area-weighted effective Eddington luminosity ($L_{\mathrm{eff}}$) and accretion luminosity on the surface of neutron star ($L_{\mathrm{acc}}$):  
\begin{subequations}
\begin{align}
    &L_{\mathrm{eff}} = \frac{A}{4\pi R_{\star}^2} L_{\mathrm{Edd}}
    \quad, 
    \\
    &L_{\mathrm{acc}} = \frac{GM_{\star}\rho_{\mathrm{acc}}A v_{\mathrm{ff}}}{R_{\mathrm{\star}}}
    \quad, 
    \\
    &\epsilon = L_{\mathrm{acc}} / L_{\mathrm{eff}}
    \quad. 
\end{align}
\end{subequations}
Here $A=4\pi R_{\star}^2 \sin^2(\theta_{\mathrm{c}}/2)$ is the transverse area of the column on the neutron star surface and $L_{\mathrm{Edd}}=4\pi GM_{\star}c/\kappa_s$ refers to the Eddington luminosity of the entire neutron star, where we adopt $\kappa_s=0.34~\mathrm{g~cm^{-2}}$ as the electron scattering opacity. Both the free-fall speed $v_{\mathrm{ff}}=\sqrt{2GM_{\star}/R_{\star}}$ and accreting inflow density $\rho_{\mathrm{acc}}$ are defined on the neutron star surface for convenience and they can be further re-scaled to set up the top boundary condition given different column heights by multiplying by factors of $(R_{\star}/r_{\mathrm{c}})^{1/2}$ and $(R_{\star}/r_{\mathrm{c}})^{3/2}$, respectively.  Therefore, the value of $\epsilon$ determines the local accretion rate and density at the top of the column.

\begin{figure}
    \centering
	\includegraphics[width=0.9\columnwidth]{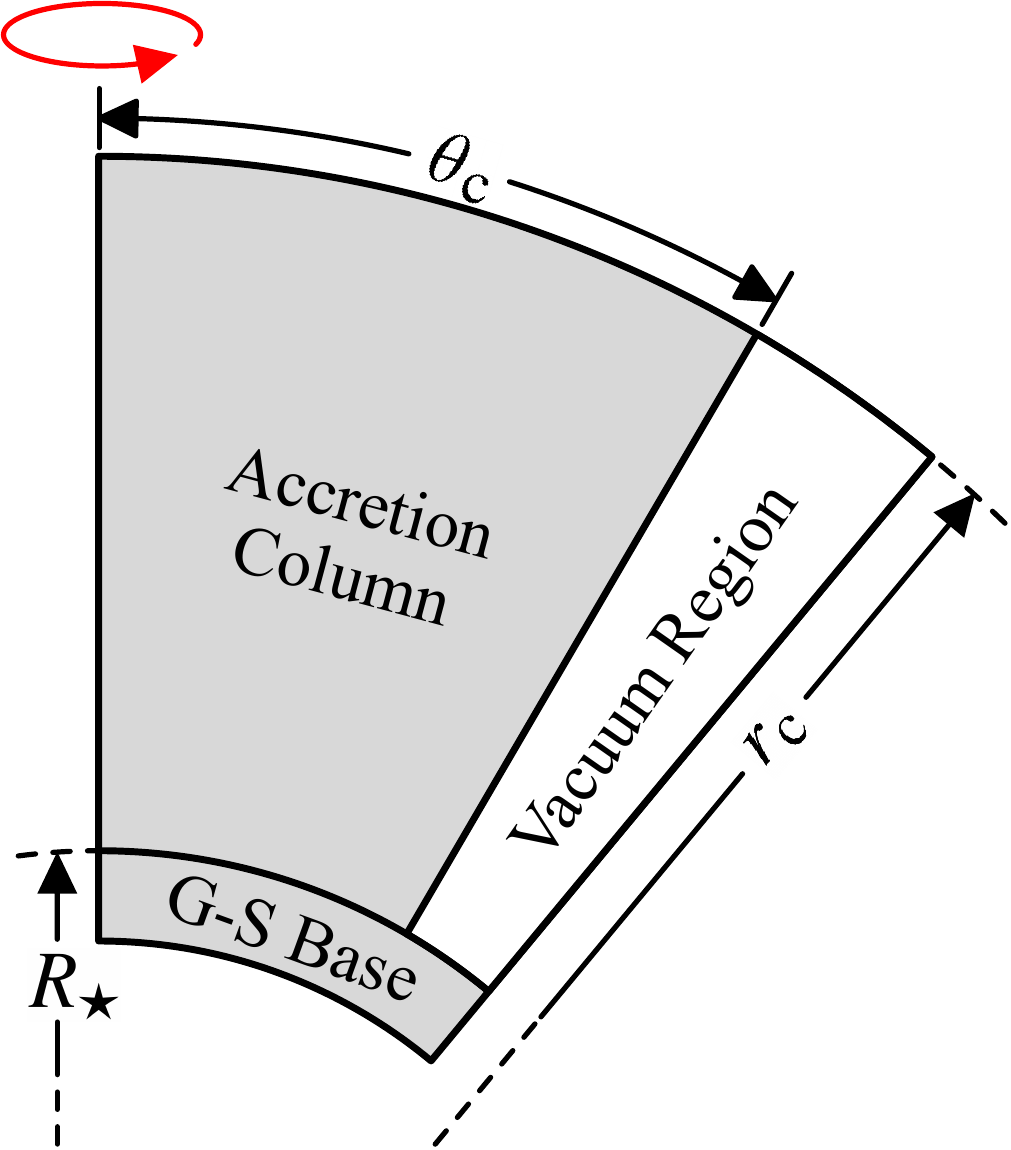}
    \caption{Geometry of the simulation domain, together with the effective boundary condition regions: the vacuum region on the outside of the column and the gas-pressure supported surface of the neutron star. The red loop is a reminder that the 2D simulation is axisymmetric.}
    \label{fig:column_partition}
\end{figure}

Similar to our previous study of columns in Cartesian geometry \citep{Paper2}, the simulation domain is partitioned into 3 different regions for distinct numerical treatments. As illustrated in \autoref{fig:column_partition}, the central region (denoted by `Accretion Column') is the main simulation region on which we focus our analyses in this paper.  The lower region (denoted by `G-S Base') and side region (denoted by `Vacuum Region') are the effective boundaries that mock up the gas pressure-supported base and vacuum region, respectively.  Our treatment of these regions will be described in detail in \autoref{sec:initial_and_boundary_conditions}. 

\subsubsection{Initial and Boundary Conditions}
\label{sec:initial_and_boundary_conditions}

\begin{table*}
	\centering
	\begin{tabular}{c l c c c c c c c c}
		\hline
		Version & \qquad Name & Mesh & $r_{\mathrm{c}}$ & $\theta_{\mathrm{c}}$ & $R_{\star}\Delta\theta$ & $\epsilon$ & $\rho_{\mathrm{acc}}$ & Accretion Rate & Total Time\\
		&  &  & ($R_{\star}$)  & & ($\mathrm{cm}$) & & ($10^{-4}~\mathrm{g~cm^{-3}}$) & ($10^{16}~\mathrm{g~s^{-1}}$) & ($t_{\mathrm{sim}}$) \\
		\hline
		\qquad\qquad\,0 (Fiducial) & Narrow-500 & $7040\times256$    & 2.5 & 0.03 & 130 & 500     & 23.00     & 12.54    & 2651    \\
		1 & Narrow-50 & $4544\times256$    & 1.8 & 0.03 & 130 & \,\;50  & \,\;2.30  & \,\;1.25 & 2651    \\
		2 & Wide-50 & $5504\times512$    & 6.0 & 0.15 & 326 & \,\;50  & \,\;2.30  & 31.28    & 3535    \\
		3 & Wide-20 & $4224\times512$    & 4.0 & 0.15 & 326 & \,\;20  & \,\;0.92  & 12.51 & 1768    \\
		4 & Narrow-500-MR & $3520\times128$    & 2.5 & 0.03 & 260 & 500     & 23.00     & 12.54    & \,\;884 \\
		5 & Narrow-500-LR & $1728\times64\,\;$ & 2.5 & 0.03 & 521 & 500     & 23.00     & 12.54    & \,\;884 \\
		\hline
	\end{tabular}
	\caption{Global parameters used to set up our six accretion column simulations. The first four simulations aim to explore the dynamical dependence on the global parameters (i.e. accretion rates and column widths), and the last two simulations are a resolution study of the fiducial simulation.  \lz{Note that what we are calling wide column simulations still have a small opening angle (0.15 radians, much less than unity).}}
	\label{tab:sim_param}
\end{table*}

Our previous work in \citet{Paper2} used an initial condition of a stationary accretion column in thermal and hydrostatic equilibrium with only vertical gradients.  Here we instead build up the accretion column using an initial condition of pure cold free-fall inflow.  The gas in the actively simulated region (`Accretion Column' in \autoref{fig:column_partition}) is initialized using the free-fall gas profile assigned with gas density $(R_{\star}/r)^{3/2}\rho_{\mathrm{acc}}$ and downward radial velocity $-(R_{\star}/r)^{1/2}v_{\mathrm{ff}}$ given a fixed accretion rate.  The gas is assumed to be cold ($T_{\mathrm{floor}}=5\times 10^6$~K, following \citealt{Klein1989}) in the beginning, so the initial gas pressure is $(R_{\star}/r)^{3/2} \rho_{\mathrm{acc}} k_{\mathrm{B}} T_{\mathrm{floor}} / m$.  The initial magnetic field is entirely radial (split-monopole) and therefore decreases with radius as $r^{-2}$.  We adopt a surface magnetic field $B_{\star}=10^{11}$~G, adequate to laterally confine the resulting column against radiation pressure.  This is below the typical field strength of high-mass X-ray binary pulsars ($\gtrsim 10^{12}$~G).  However, such strong magnetic fields will not affect the column dynamics, and they exacerbate the problem of variable inversion from conservative variables to primitive variables,
in particular in resolving the small gas internal energy (see detailed discussion in section 4.1 of \citealt{Paper2}).  The radiation field of the initial inflow is assumed to be isotropic and in local thermal equilibrium (LTE) with the cold gas in the fluid rest frame. 

We adopt a similar strategy of setting boundary conditions as in \citet{Paper2}:  effective boundaries (hereafter, soft boundaries) are designed to mock up the neutron star surface (`G-S Base' in \autoref{fig:column_partition}) at the bottom and the side vacuum region outside the accretion column (`Vacuum Region' in \autoref{fig:column_partition}).  The actual boundaries of the simulation domain (hereafter, hard boundaries) are applied directly at the top and polar axis boundaries of the active accretion column (`Accretion Column' in \autoref{fig:column_partition}) and at the exterior edges of the soft boundary regions.  Detailed justification for the soft boundary arrangements can be found in \citet{Paper2}, but we briefly review them here.  

The gas-supported base is the layer below the active region of the accretion column and is initialized in hydrostatic equilibrium with an isothermal non-degenerate cold gas ($T_{\mathrm{b}}=5\times 10^6$~K). We adopt a neutron star surface density $\rho_{\star}=10^6~\mathrm{g~cm^{-3}}$ to calculate the gas density profile at the base $\rho_{\mathrm{b}}=\rho_{\star}\exp{(-r/h_\mathrm{b})}$, where $h_\mathrm{b}=R_{\star}T_{\mathrm{b}}/g_{\mathrm{b}}$ is the scale height.  We set the effective gravitational acceleration $g_{\mathrm{b}}$ to be smaller than the surface gravitational acceleration $g_{\star}=GM_{\star}/R_{\star}^2$ by at least a factor of 500 in our simulations so that a scale height can be resolved by at least 5 grid cells to numerically maintain the hydrostatic equilibrium. The initial magnetic and radiation intensity fields are set to be the same as the active accretion column region.  The height of the gas-supported base is set to 6 scale heights for each simulation.  

The advantage of using this effective boundary is that the magnetic confinement at the bottom of the accretion column can be automatically achieved without directly implementing any cumbersome boundary condition that invokes artificial magnetic tension and pressure forces.  However, as the accretion column evolves, the gas-supported base will be continuously heated, causing radiation diffusion, and radiation pressure forces, towards the side.  Although the tendency to horizontally bend magnetic field is small because of the huge gas inertia, this tendency is finite and can keep accumulating.  As long as the simulation runs sufficiently long, this will eventually tilt the magnetic field away from the radial direction which will then fail to confine the accretion column above.  To alleviate this numerical issue, we adopt an exponential damping of the horizontal velocity in the base on a sound crossing time across a grid cell ($t_{\mathrm{s}}$) using $d\ln{v_\theta}=-dt/t_{\mathrm{s}}$, where $v_\theta$ refers to the instantaneous horizontal speed.  

The vacuum effective boundary at the side is an optically thin region where the density and gas pressure are both set to the floor values at all times but the velocity and magnetic fields are allowed to evolve.  This soft side boundary is necessary to simultaneously achieve both the transverse magnetic confinement and the free escape of photons at the side.  The width of the vacuum region is set to be 10\% of the domain size.  The initial velocity in this region is set to be 0. The magnetic and radiation intensity fields are initially set the same as in the active accretion column region.  More details about this effective boundary condition can be found in section 2.4 of \citet{Paper2}.

The hard boundary conditions of the simulation domain are similar to \citet{Paper2} except the central boundary, where the polar wedge boundary condition \citep{Stone2020} is adopted for both gas and radiation.  This enforces the axisymmetric constraint of no velocity, magnetic field, or radiation flow through the pole.  Note that the computation in spherical polar coordinates can have numerical problems from the geometric terms when the polar region is highly resolved.  For example, the problem can arise from the subtraction of similar numbers in the geometric terms if the interval between adjacent polar angles is too small.  This can lead to inaccurate outcomes when computing the surface flux from the variable reconstruction at the cell face near the polar region.  Therefore, we lower the order of the reconstruction near the polar region to avoid this numerical effect.

The bottom hard boundary condition of the simulation domain is reflective for both gas and radiation, and enforces the same magnetic field as the initial condition.  The outer side boundary conditions of the gas and magnetic field are set to be reflective as well but the radiation uses a vacuum boundary condition so that it can escape \citep{Paper1}.  At the top, the cold gas ($5\times 10^6$~K) is set to free fall from the boundary within the active accretion column region with its density $(R_{\star}/r_{\mathrm{c}})^{3/2} \rho_{\mathrm{acc}}$ at downward radial velocity $-(R_{\star}/r_{\mathrm{c}})^{1/2} v_{\mathrm{ff}}$, while the radiation intensity field is set to be isotropic and in LTE with the gas in the comoving frame.  Outside the accretion column region, the top boundary condition above the vacuum region is simply set to be outflow for the gas \citep{Paper1} and vacuum for the radiation. The magnetic field at the top of the domain is fixed as in the initial condition.

\subsubsection{Simulation Parameters}
\label{sec:simulation_parameters}

We simulated four accretion columns in a split-monopole magnetic field with various widths and accretion rates, plus an extra two for a resolution study.  Parameters for these simulations are listed in \autoref{tab:sim_param}.  We take Version~0 as our fiducial simulation and set up Version~1 by decreasing the accretion rate by a factor of 10.  We also set up two accretion column simulations $3\times$ wider than the fiducial column, where the first one (Version~2) has the same $\epsilon$ and the second one (Version~3) decreases the accretion rate by a factor of 2.5. We lower the resolution by a factor of 2.5 for the two wide column simulations due to the high computational cost. Besides these four simulations exploring the parameter space of column width and accretion rate, we also conduct a resolution study by decreasing the resolution by factor of 2 and 4 with respect to the fiducial version.  We perform our analysis of the simulation output data dumps with the time resolution $t_{\mathrm{sim}}=2.8\times10^{-6}$~s.

\section{Results}
\label{sec:results}

\begin{figure*}
    \centering
	\includegraphics[width=\textwidth]{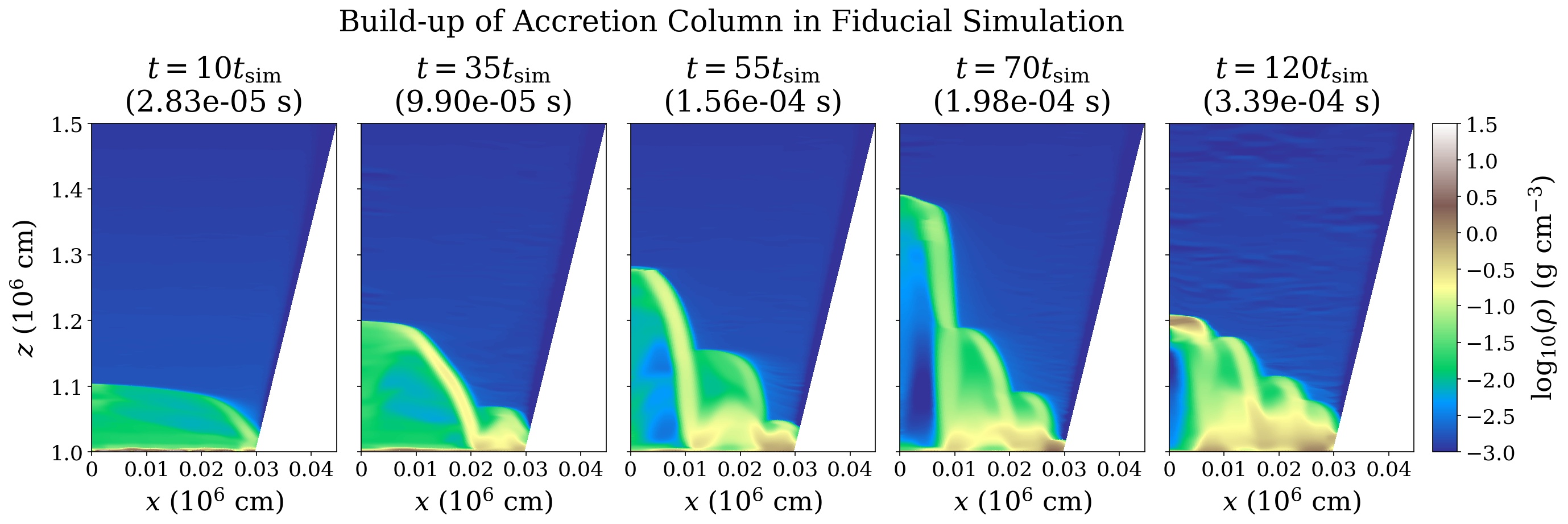}
    \caption{2D density profiles during the column build-up epoch of the fiducial simulation.  These five snapshots illustrate how the column is built up from the cold accretion flow.  After the column structure is established at $\simeq70t_{\mathrm{sim}}$, the accretion column soon enters into a long-lived, oscillatory phase. 
    }
    \label{fig:build_up_fiducial}
\end{figure*}

We focus our discussion here on the results of our fiducial simulation, which shares many properties with the other simulations.  We comment where necessary how the global parameters of column width and accretion rate affect the results of our other simulations.  In \autoref{sec:build_up_of_accretion_column}, we describe
how the accretion column is built up by forming a sinking zone under the accretion shock.  In \autoref{sec:high_frequency_oscillatory_behavior}, we discuss the oscillatory behavior of the accretion column in detail.  In \autoref{sec:luminosity_variation}, we present the light curves of all the simulations and discuss how the shock oscillation affects the emergent radiation by showing their variability power spectra.  In \autoref{sec:time_averaged_profiles}, we compare the time-averaged profiles with the 1D stationary model incorporating different shapes of the sinking zone.  In \autoref{sec:energetics_of_accretion_columns}, we examine the energetics of the accretion column, and show how radial advection competes with \textit{pdV} work in providing thermal pressure support of the accretion column.  In \autoref{sec:entropy_wave_and_resolution_study}, we identify the entropy wave existing in the sinking zone and conduct a resolution study of the fiducial simulation.  A set of movies of various quantities in our four high-resolution simulations is available online at this YouTube \href{https://youtube.com/playlist?list=PLbQOoEY0CFpWubAItXtHIIRdGhlBuftch}{link}\footnote{\url{https://youtube.com/playlist?list=PLbQOoEY0CFpWubAItXtHIIRdGhlBuftch}}.  

\subsection{Build-up of Column}
\label{sec:build_up_of_accretion_column}

All the simulations are initialized with only the cold gas free-falling onto the neutron star, where the gas can reach speeds at the stellar surface of up to $\sim0.64c$ ($\Gamma=1.31$).  This gas is immediately halted by the rigid neutron star surface and forms a shock that releases much of its kinetic energy into radiation.  A portion of the radiation directly escapes sideways, and the rest propagates upward and interacts with the incoming accretion flow.  Note that radiation diffusion into the neutron star is negligible because the diffusion time scale below the stellar surface is much longer than the cooling time scale of the column itself.  At the high accretion rates we adopt here, the radiation pressure support exceeds the ram pressure and the gravity of the downward gas flow and therefore pushes the shock away from the stellar surface as shown in the left four panels of \autoref{fig:build_up_fiducial}.  The shock front keeps rising until an equilibrium is roughly achieved among radiation pressure support, ram pressure, and gravity.  Hence, a radiation pressure dominated region below the shock front is naturally formed as \citet{Basko1976} predicted, which we call the `sinking zone'.
\lz{This region is thoroughly confined by the magnetic field, with the highest plasma beta (thermal pressure/magnetic pressure) occurring at the base of the sinking zone, and ranging from 0.03 to 0.08 for the different simulations. }

While the shock is elevating and the sinking zone is being established, a wave is generated at the outer part of the column and then propagates towards the center as shown beginning at $t=35t_{\mathrm{sim}}$ (the second panel in \autoref{fig:build_up_fiducial}).  As we discuss below in \autoref{sec:entropy_wave_and_resolution_study}, this is the entropy wave associated with the slow-diffusion photon bubble instability \citep{Arons1992, Paper1}.  This entropy wave is periodically excited as shown in the snapshot at $t=55t_{\mathrm{sim}}$ (the third panel in \autoref{fig:build_up_fiducial}), and a lateral structure consisting of radial fingers forms that resembles what we observed in the Cartesian column simulations (see version~2 in \citealt{Paper2}).  The height of each `finger' increases towards the center of the column and these `fingers' build up a mound-shaped column structure.  The generation of these entropy waves persists through all our simulations, propagating from the outermost parts of the column inward to the polar axis.  \lz{Recall that the gas is magnetically constrained to move only radially along the field lines as the finger-shaped structures oscillate up and down.  Therefore, the wave pattern that propagates inwards is entirely driven by the phase differences of the oscillations at different transverse locations.}  As the waves reach the polar axis, they merge with the finger-shaped structure there that oscillates up and down (see $t=70t_{\mathrm{sim}}$ and $t=120t_{\mathrm{sim}}$ in \autoref{fig:build_up_fiducial}).

\subsection{High-Frequency Oscillatory Behavior}
\label{sec:high_frequency_oscillatory_behavior}

After the accretion column finishes the build-up of the sinking zone, it starts to oscillate at high frequency ($\gtrsim$kHz) and this oscillation persists through the end of the simulation.  The physical origin of this oscillation is similar to what we found in our previous Cartesian simulations: it is caused by the instantaneous imbalance between global heating and cooling due to inefficient radial energy transport.  Indeed, if we estimate twice the cooling time using the same method as in section~3.3 of \citealt{Paper2} \lz{(i.e. assuming that the contraction and expansion of the accretion column take approximately the same amount of time)}, we find a value of $\sim 0.5~\mathrm{\mu s}$, comparable to the period measured from the simulation ($\sim 0.2~\mathrm{\mu s}$).  Note that we do not see any evidence of the formation of pre-shocks in our split monopole simulations, in contrast to the Cartesian simulations of \citet{Paper2} where pre-shocks formed in the free-fall region and then fell onto the sinking zone. 

\begin{figure*}
    \centering
	\includegraphics[width=\textwidth]{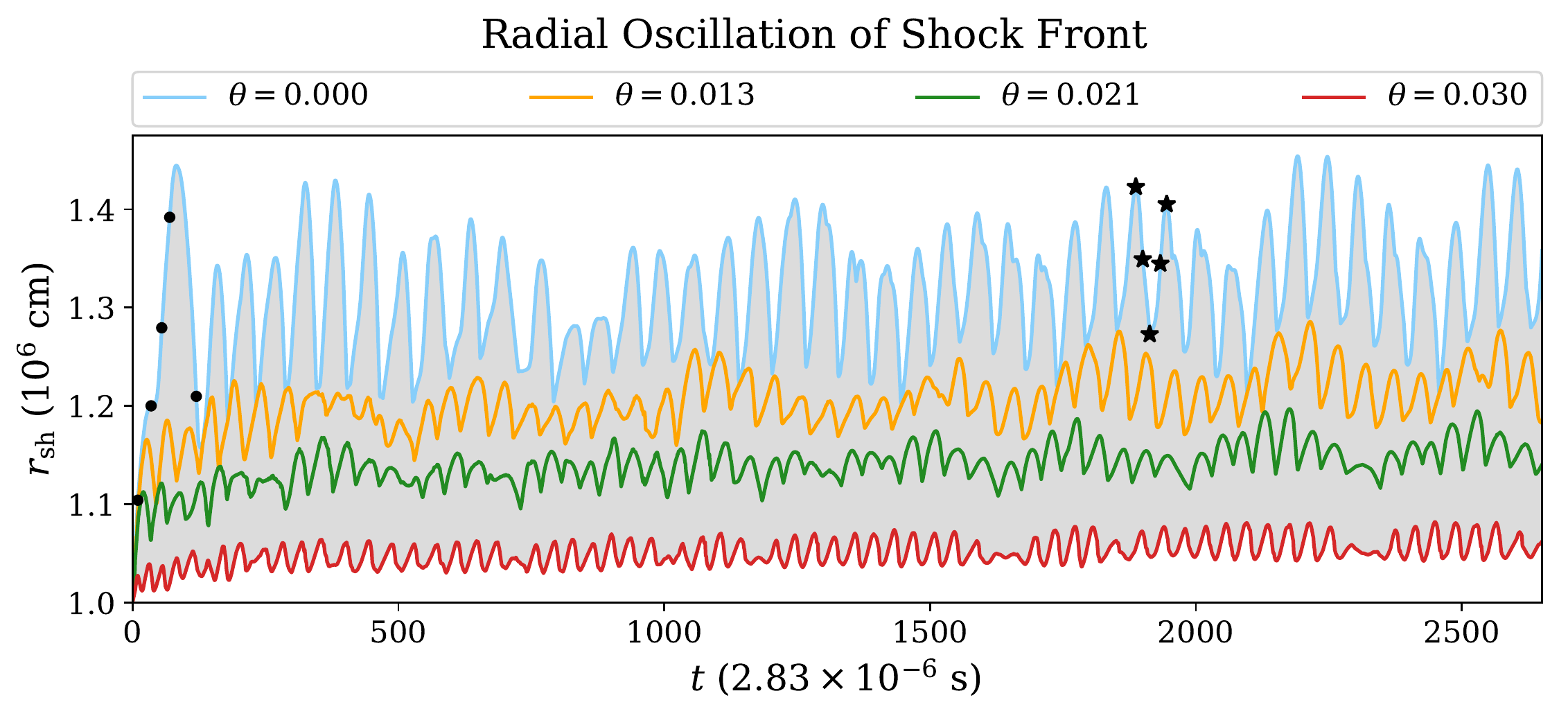}
    \caption{Oscillation of the shock fronts at different transverse positions in the fiducial simulation.  The shock oscillation amplitude increases, and the frequency decreases, towards the polar axis of the column.  The shaded region indicates the range of the shock oscillation amplitude from the polar axis to the outer edge of the accretion column.  The five black dots refer to the snapshots in \autoref{fig:build_up_fiducial} and the five black stars represent the snapshots in \autoref{fig:oscillation_fiducial}.  }
    \label{fig:shock_oscillation_fiducial}
\end{figure*}

\begin{figure*}
    \centering
	\includegraphics[width=\textwidth]{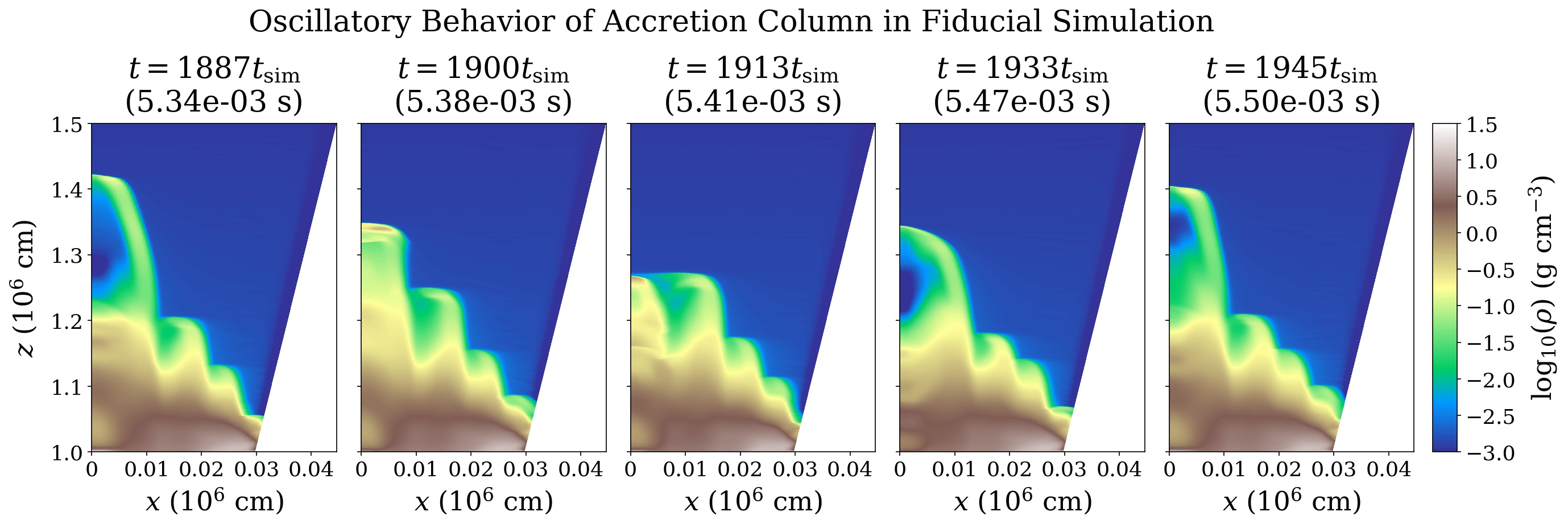}
    \caption{2D density profiles over one polar axis oscillation period of the accretion column for the fiducial simulation.  
    }
    \label{fig:oscillation_fiducial}
\end{figure*}

Both the oscillation amplitude and frequency show considerable variation with transverse location in the column.  This is evident from \autoref{fig:shock_oscillation_fiducial}, which shows the shock height ($r_{\mathrm{sh}}$) at different transverse positions as a function of time.  Note that the shock oscillations are quite coherent at any particular transverse location.  The oscillation amplitude is highest, and the frequency is lowest, at the center (polar axis) of the column.  In \autoref{fig:oscillation_fiducial}, we present 2D density profiles over a time span equal to the oscillation period at the center of the column in the fiducial simulation.  The physical reason for the transverse variation in oscillation frequency is that the closer one is to the polar axis, the longer it takes for radiation to diffuse outward in the transverse direction.  This implies a longer time scale for the system to respond to heating or cooling closer to the polar axis.  

\begin{figure*}
    \centering
	\includegraphics[width=0.795\textwidth]{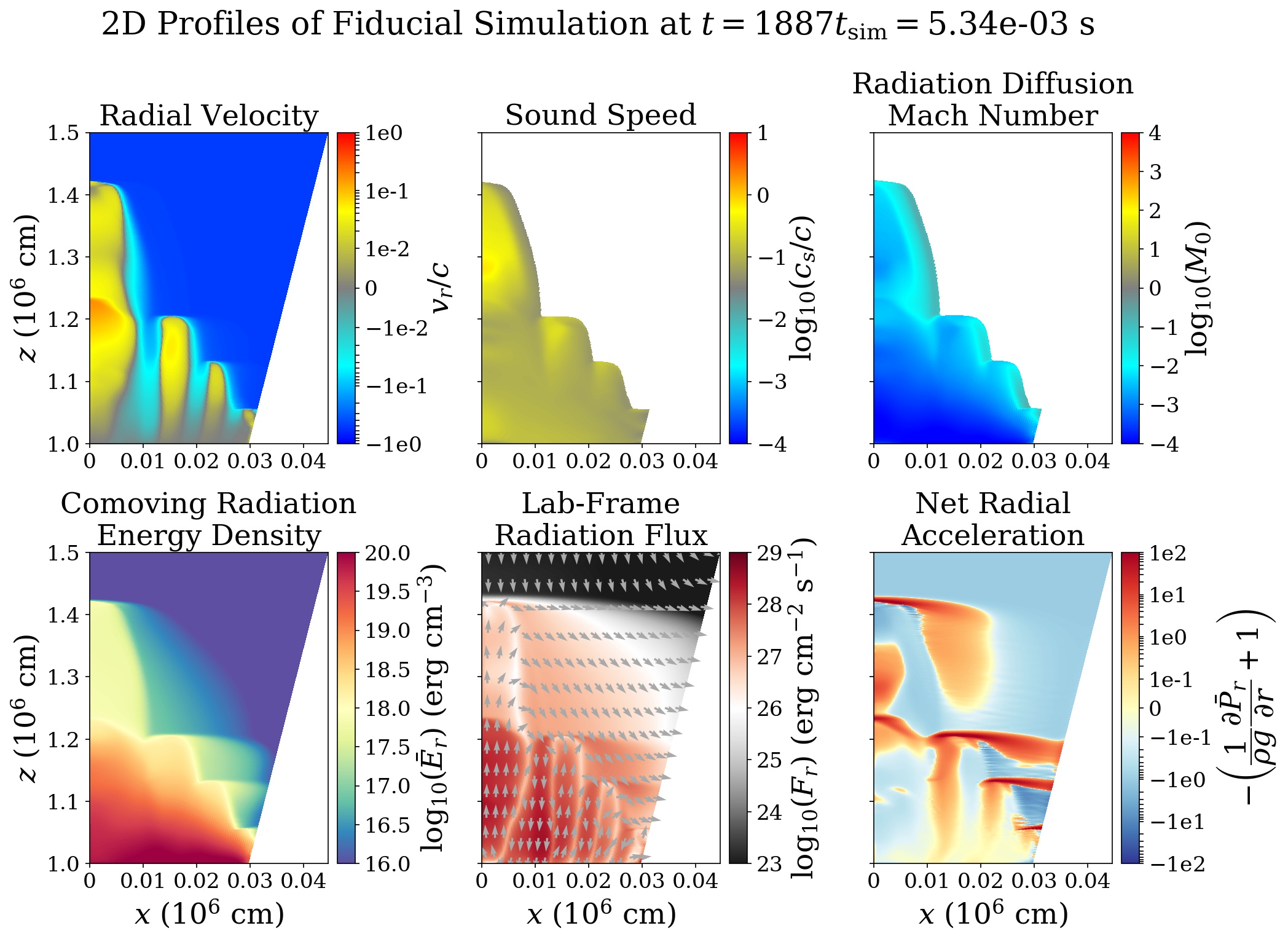}
    \caption{2D profiles of six quantities in the fiducial simulation when the accretion column is mostly extended at $t=1887t_{\mathrm{sim}}$ (see the first panel of \autoref{fig:oscillation_fiducial}).  }
    \label{fig:snapshot1_fiducial}
\end{figure*}

\begin{figure*}
    \centering
	\includegraphics[width=0.795\textwidth]{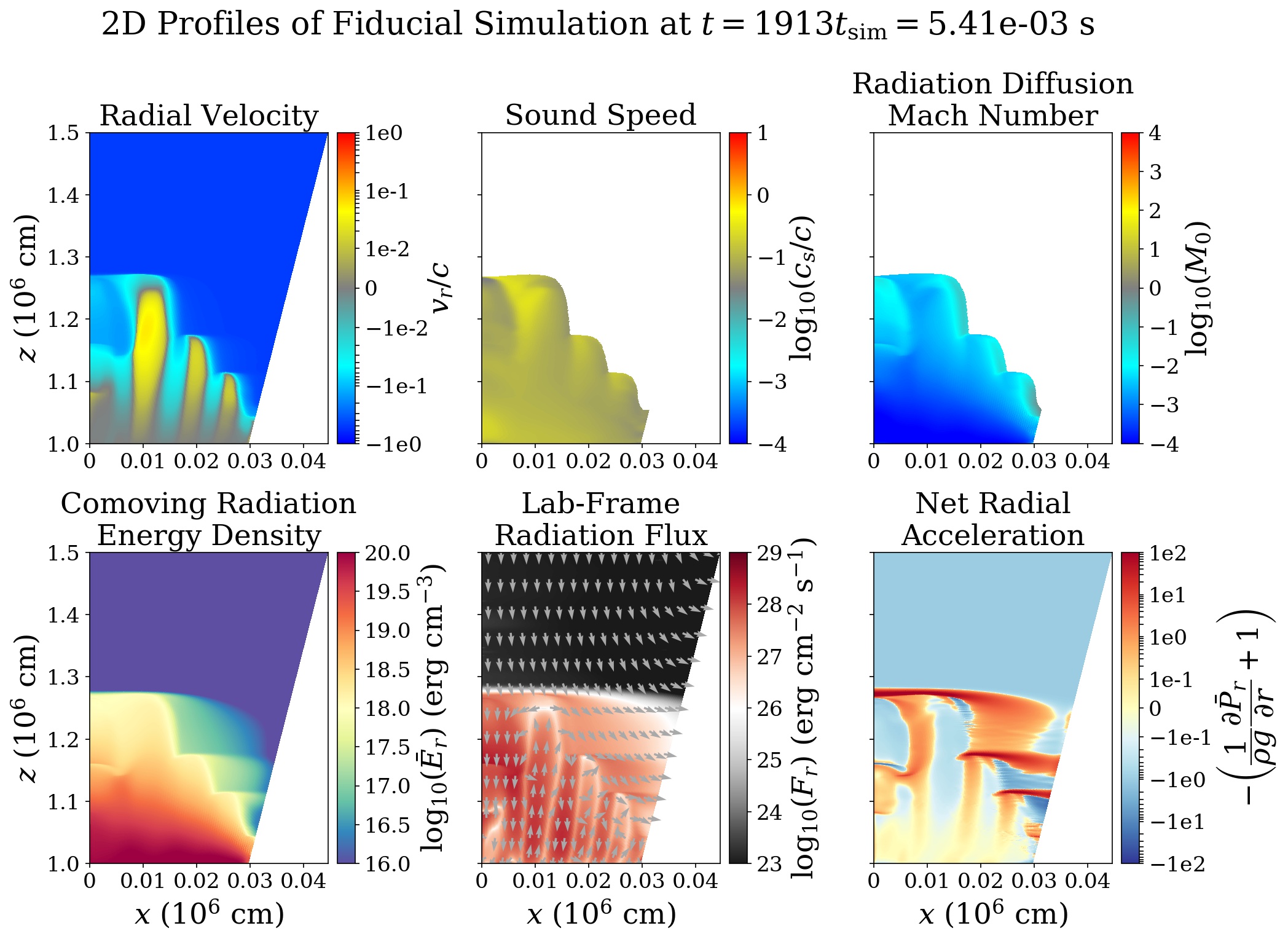}
    \caption{2D profiles of six quantities in the fiducial simulation when the accretion column is mostly compressed at $t=1913t_{\mathrm{sim}}$ (see the middle panel of \autoref{fig:oscillation_fiducial}).}
    \label{fig:snapshot2_fiducial}
\end{figure*}

\begin{figure*}
    \centering
	\includegraphics[width=\textwidth]{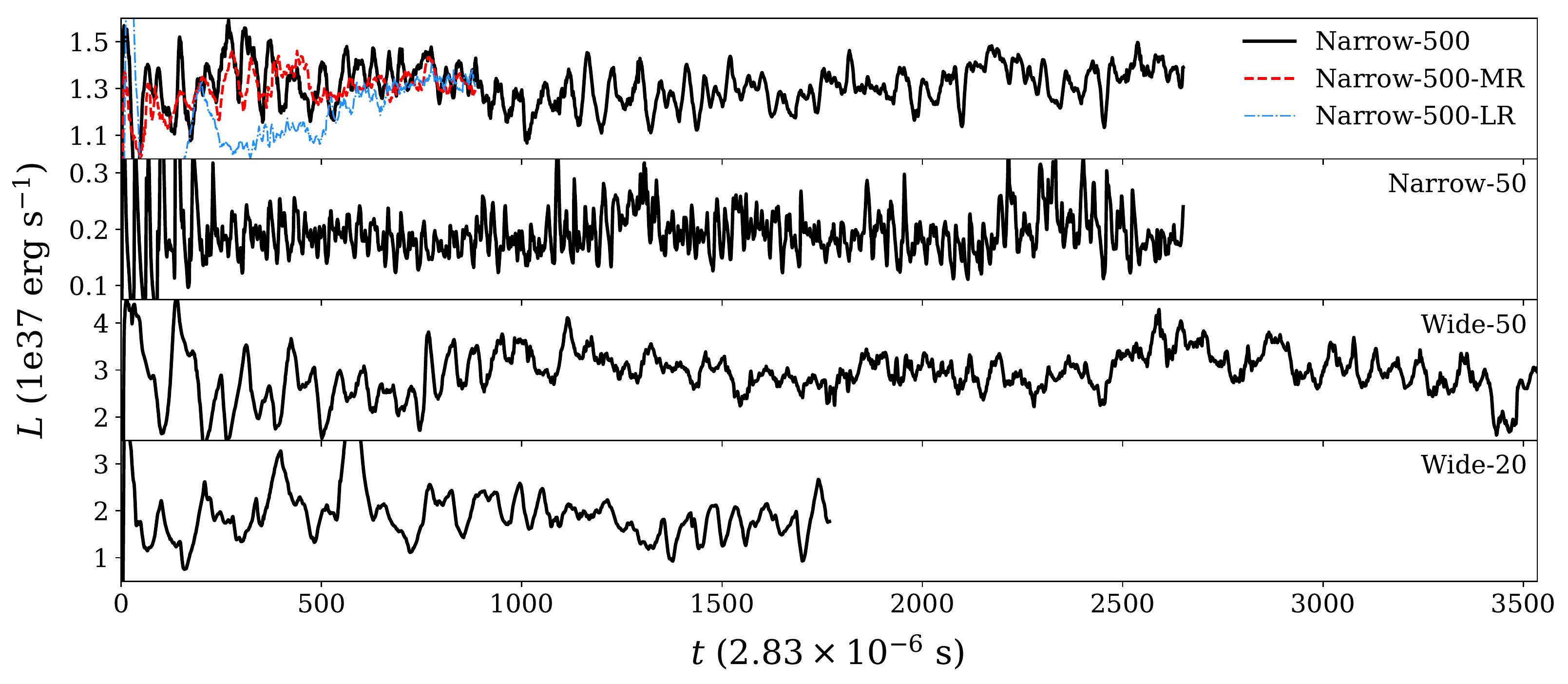}
    \caption{Light curves of all six accretion column simulations which have different accretion rates, different column widths, and/or different resolution.}
    \label{fig:lightcurve}
\end{figure*}

\begin{table*}
	\centering
	\begin{tabular}{c l c c c c c c}
		\hline
		Version & \qquad Name & $\left<L\right>$ & $\left<L_{\mathrm{iso}}\right>$ & $\sigma(L)/\left<L\right>$ & $f_{\mathrm{peak}}$ & $f_{\mathrm{break}}$ \\
		&& ($10^{37}\ \mathrm{erg~s^{-1}}$) & ($10^{39}\ \mathrm{erg~s^{-1}}$) & (\%) & (kHz) & (kHz) \\
		\hline
		\qquad\qquad\,0 (Fiducial) & Narrow-500 & 1.31 & 3.27 & 6.28 & 5.86 & 9.66 \\
		1 & Narrow-50 & 0.19 & 0.57 & 19.28 & 10.35 & 15.87 \\
		2 & Wide-50 & 3.02 & 0.30 & 12.67 & 2.00 & 6.19 \\
		3 & Wide-20 & 1.85 & 0.11 & 19.74 & 3.45 & 6.55 \\
		4 & Narrow-500-MR & 1.32 & 4.09 & 3.90 & 6.08 & 11.05 \\
		5 & Narrow-500-LR & 1.31 & 4.41 & 2.92 & 6.89 & 22.06 \\
		\hline
	\end{tabular}
	\caption{Parameters of the light curves for all six simulations.}
	\label{tab:qpo_characteristic}
\end{table*}

In \autoref{fig:snapshot1_fiducial} and \autoref{fig:snapshot2_fiducial}, we present 2D profiles of other quantities at the maximum ($t=1887t_{\mathrm{sim}}$) and minimum ($t=1913t_{\mathrm{sim}}$) heights of the oscillation at the polar axis.  The finger-shaped pattern is prominent in the radial velocity ($v_r$, see the upper left panels in \autoref{fig:snapshot1_fiducial} and \autoref{fig:snapshot2_fiducial}), with the gas in adjacent finger-shaped regions moving in opposite directions.  As we discuss in detail in \autoref{sec:entropy_wave_and_resolution_study}, this propagating wave pattern toward the polar axis is a manifestation of the entropy modes that are associated with the slow-diffusion photon bubble instability.

The sound speed ($c_s$) is dominated by radiation pressure and is mildly relativistic ($\sim0.1c$, see upper second panels in \autoref{fig:snapshot1_fiducial} and \autoref{fig:snapshot2_fiducial}) in the sinking zone during the oscillation.  This is in general faster than the magnitude of the gas velocity ($\sim0.01c$), so that the oscillations are nearly hydrostatic, except during outward moving phases along the polar axis where the radial speed can slightly exceed the sound speed.  This agrees with the fact that the net radial acceleration is only tens of percents of the local gravitational acceleration throughout most of the sinking zone (see the lower right panels in \autoref{fig:snapshot1_fiducial} and \autoref{fig:snapshot2_fiducial}).  The radiation diffusion Mach number ($M_0$) refers to the ratio of the radiation diffusion speed $c/\tau$ (where $\tau$ is the optical depth over a pressure scale height) to the sound speed (for details see equation~A14 in \citealt{Paper1}). As shown in the upper third panels in \autoref{fig:snapshot1_fiducial} and \autoref{fig:snapshot2_fiducial}, $M_0$ is small everywhere in the sinking zone, indicating that the radiation diffuses much more slowly ($\sim10^{-4}c$) compared with the sound speed (as well as the gas velocity).  Hence, the sinking zone is always in the slow-diffusion regime during the oscillation. 

During the oscillation, the radial component of lab-frame radiation flux ($F_r$) is dominated by the advection and therefore its direction is the same as the gas radial velocity, while the transverse component indicates the sideways cooling process (see the lower middle panels in \autoref{fig:snapshot1_fiducial} and \autoref{fig:snapshot2_fiducial}).  When the finger-shaped structure is most elongated, cooling of the accretion column becomes most efficient due to the maximum sideways cooling area. Global cooling exceeds heating at this time, and the column therefore loses radiation pressure support against gravity and thus collapses.  When the column is at its minimum height, the sideways cooling area of the sinking zone is the smallest, and heating then exceeds the sideways cooling, leading to the column structure expanding radially outward again.  Because the over-cooling of the innermost region is usually stronger at maximum radial extent than the over-heating at minimum radial extent, the innermost finger structure of the accretion column always collapses faster than its expansion.

\subsection{Luminosity Variation}
\label{sec:luminosity_variation}

\begin{figure}
    \centering
	\includegraphics[width=\columnwidth]{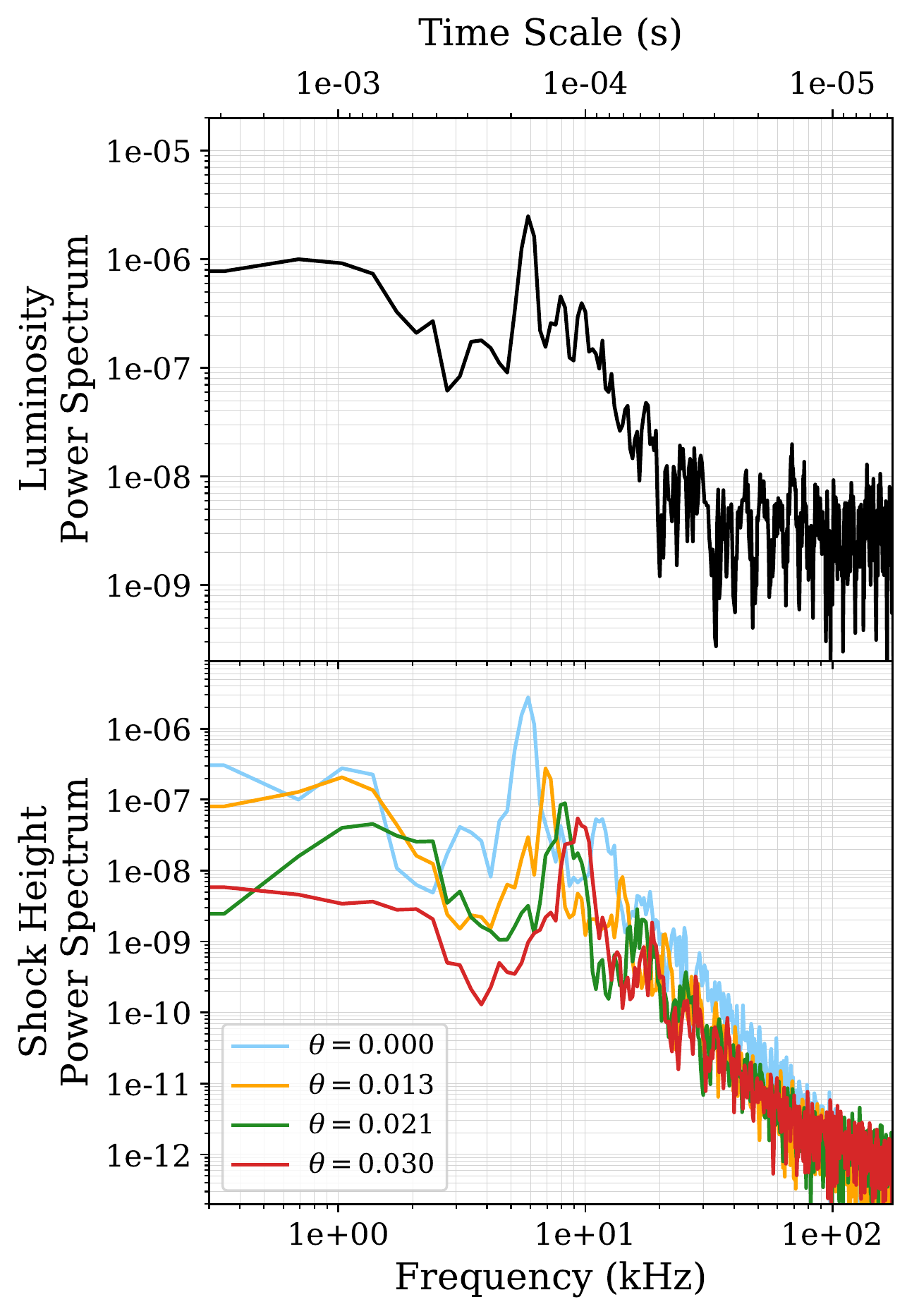}
    \caption{Power spectra of the light curve and shock oscillation height at various transverse locations in the fiducial simulation.  The power spectra are calculated using Welch's method with a Hann window function, a segment length of 2.9~ms, and a sampling interval of $t_{\mathrm{sim}}$.  These two plots suggest that the light variation arises from the shock oscillations, where the light curve is dominated by the shock oscillation in the innermost region.} 
    \label{fig:power_spectrum_fiducial}
\end{figure}

The oscillatory behavior of the accretion column produces variability in the emergent radiation in all our simulations, as illustrated in \autoref{fig:lightcurve}.  Various parameters measured from these light curves are listed in \autoref{tab:qpo_characteristic}.  These include the time-averaged luminosity $\left<L\right>$ as well as the time-averaged apparent luminosity  $\left<L_{\mathrm{iso}}\right>$ that an observer might infer \lz{from seeing only one accretion column at one magnetic pole}.  For the latter we measured the side area of the cone enclosing the edge of the accretion column, up to a height that captures 99 percent of the emitted luminosity in the time-average.  We then enhanced $\left<L\right>$ by the ratio of the surface area of the neutron star to this area.  \autoref{tab:qpo_characteristic} also lists the standard deviation of the luminosity $\sigma(L)$ scaled with that mean, which is positively correlated across our simulations with the oscillation amplitude of the accretion shock on the polar axis. 

All of our light curves exhibit quasi-periodic oscillations (QPOs), and we include their frequencies $f_{\mathrm{peak}}$ in \autoref{tab:qpo_characteristic}.  These frequencies range from $\sim$2 to 10~kHz, depending on the simulation parameters.  Increasing the accretion rate at fixed column width generally results in smaller fractional variability and lower QPO frequencies (compare versions 0 to 1, and versions 2 to 3).  As we discussed in \citet{Paper2}, the oscillations are driven by a mismatch between the ability to distribute accretion energy radially in the column and the cooling at each height.  Higher accretion rate generally results in a taller column so that it takes longer for radial advection in the sinking flow to distribute energy from the shock downward, resulting in lower frequencies in the oscillation.  As we discuss in more detail below in \autoref{sec:energetics_of_accretion_columns}, \textit{pdV} work by the sinking flow also becomes more important and provides a local source of energy to balance the cooling, so that the oscillation amplitude is reduced.  Increasing the column width at fixed effective Eddington ratio reduces the fractional variability and decreases the QPO frequency (compare versions 1 and 2).  Because the transverse diffusion time is longer in a wider column, cooling is less efficient, partially alleviating the mismatch and reducing the amplitude of the oscillation.  The longer cooling time also results in a longer oscillation period. 

\autoref{fig:power_spectrum_fiducial} shows a power spectrum of the light curve of the fiducial simulation, as well as a power spectrum of the shock height at various transverse locations.  The QPO in the luminosity power spectrum consists of a peak together with an asymmetric shoulder extending to a higher frequency $f_{\mathrm{break}}$.  All our simulations exhibit this characteristic QPO shape, and we also list the values of $f_{\mathrm{break}}$ in \autoref{tab:qpo_characteristic}.  The reason for this breadth in frequency space is that the shock height oscillates with different amplitudes and frequencies depending on the transverse location within the accretion column (recall \autoref{fig:shock_oscillation_fiducial}).  The shock height at the polar axis center of the column oscillates with the highest amplitude and lowest frequency, and this produces the main peak $f_{\mathrm{peak}}$ in the light curve power spectrum, particularly as the high amplitude results in a large oscillating sideways emitting area.  The higher frequency, smaller amplitude shock oscillations away from the polar axis contribute to the high frequency shoulder in the light curve power spectrum, with $f_{\mathrm{break}}$ corresponding to the shock oscillation frequency at the outer edge of the column.  As is evident from \autoref{tab:qpo_characteristic}, wider columns show larger values of $f_{\mathrm{break}}/f_{\mathrm{peak}}$, and therefore broader high frequency shoulders in log frequency space.

\subsection{Time-averaged Profiles}
\label{sec:time_averaged_profiles}

\begin{figure*}
    \centering
	\includegraphics[width=\textwidth]{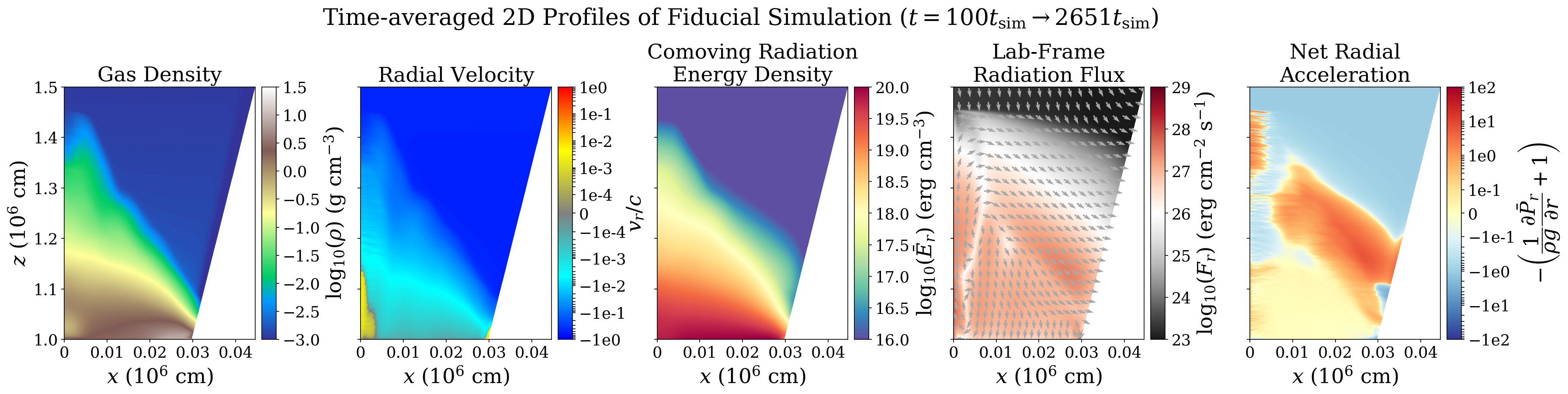}
    \caption{Time-averaged 2D profiles of density, mass-weighted radial velocity, comoving energy density, lab frame radiation flux, and mass-weighted radial acceleration in the fiducial simulation.  The shock structure and entropy wave are smeared out by the time average. 
    }
    \label{fig:timeavg2D}
\end{figure*}

\begin{figure*}
    \centering
	\includegraphics[width=0.75\textwidth]{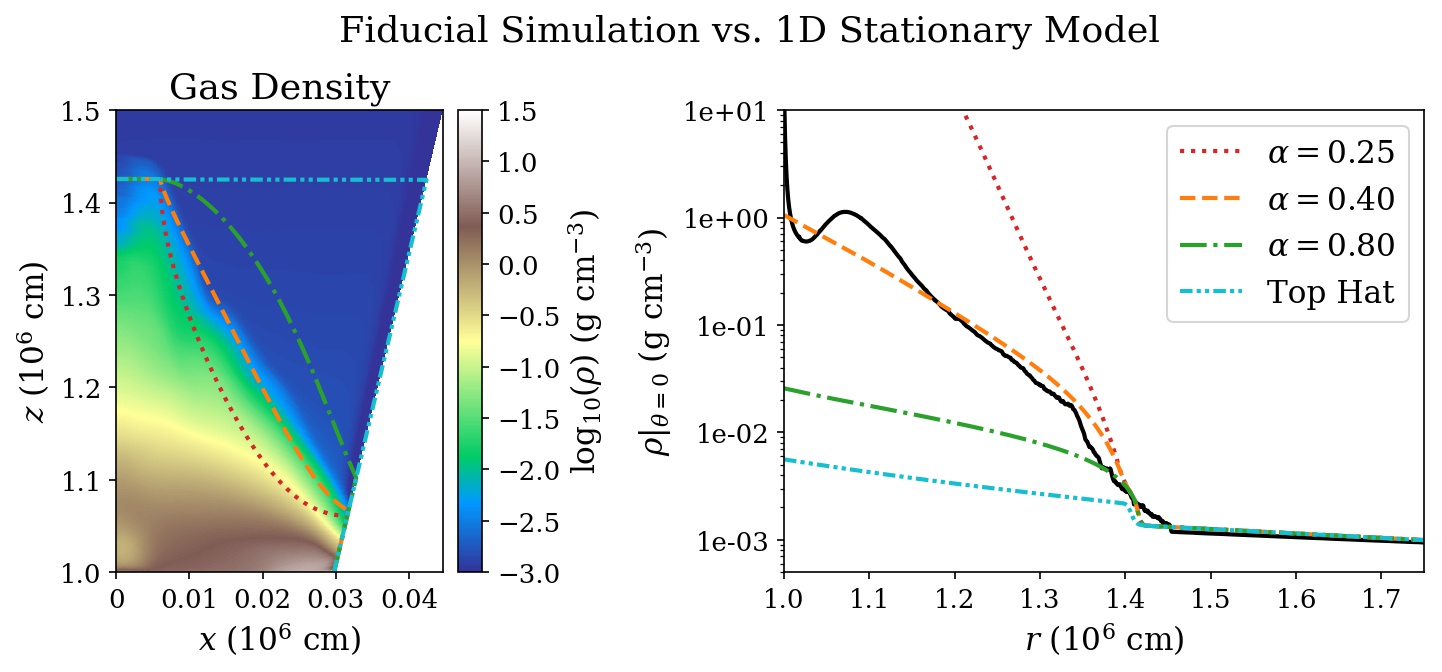}
    \caption{Comparison between the time-averaged density profile and the 1D stationary model using different shapes of the sinking zone.  In the left panel, the curves in various line styles and colors represent different assumed shapes for the sinking-zone integration.  In the right panel, the black solid curve is the simulation data, while other curves represent the integrated results from the 1D stationary model using the corresponding shape functions as shown in the left panel.  The $\alpha=0.40$ shape function is closest to the simulated 2D shape in the left panel, consistent with the fact that the corresponding 1D stationary solution is closest to the radial profile of the simulation data along the polar axis in the right panel. }
    \label{fig:timeavg1D_compare}
\end{figure*}

The accretion column is highly dynamical, but its short oscillation time scale ($\lesssim$1~ms) may make it challenging to detect with existing observational facilities.  Virtually all observational studies of accreting neutron stars average over longer time scales.  Here we therefore time-average our simulations in order to make contact with such observations.  \autoref{fig:timeavg2D} shows 2D time-averaged profiles of various quantities in the fiducial simulation.  The jump in density and other quantities associated with the oscillating shock front has been smoothed out, as has radial finger structures associated with the propagating entropy waves. The latter is in contrast to what we found in the time-averaged profiles of the Cartesian simulations (see Fig.~15 of \citet{Paper2}, where the pre-shock disturbed the entropy wave in the sinking zone, causing it to leave residual vertical striations in the time average. 

As shown in the first and third panels of \autoref{fig:timeavg2D}, the gas density and the radiation internal energy in the sinking zone increase exponentially inward towards the stellar surface. Despite the fact that the gas below the shock front oscillates along the magnetic field lines, the second panel of \autoref{fig:timeavg2D} shows that the time-averaged, mass-weighted velocity away from the polar axis is small and downward, consistent with the physical picture of \citet{Basko1976} and with the overall sinking flow of accreting matter. Note, however, that gas in the lower part of the central polar axis region is moving outward in the time average, because the collapse is faster than the expansion of the innermost finger structure.  The sinking zone achieves approximate hydrostatic equilibrium in its time average, as indicated by the nearly zero mass-weighted radial acceleration in the last panel of \autoref{fig:timeavg2D}. Even though the comoving radiation flux must be upward in the sinking zone to provide support against gravity, the lab-frame radiation flux away from the polar axis is still downward, as shown in the second to last panel of \autoref{fig:timeavg2D}. Hence, the downward advection must be dominant over the upward radial diffusion and this transports the heat from the shock front to the bottom of the column.

In \citet{Paper2}, we found that the time-averaged mound shape of the shock front has a higher cooling efficiency compared to the assumed top-hat geometry in the one-zone model of \citet{Basko1976}.  Here, we attempt a comparison of our 2D simulation results with the approximate 1D stationary solution.  We do this by numerically integrating the column profiles at the polar axis $\theta=0$ starting above the shock and going inward to the stellar surface by using the 1D stationary model equations.  We adopt the time-averaged radial profile from the simulation in the free-fall zone, and then integrate the 1D equations inward starting from $r=1.5 R_\star$ (the point at which the density starts to increase because of the smoothed-out, oscillating shock).  In order to explore how the mound shape geometry influences the trend of the profiles, we also introduce a shape function (equation~\ref{eq:mod_bump_function}) that approximately accounts for the mound shape of the shock front.  The details of the formulation can be found in \hyperref[sec:appendix_1]{Appendix A}.

We adopt four different shape functions, which are shown on top of the time-averaged density in the fiducial simulation in \autoref{fig:timeavg1D_compare}.  The cyan dot-dot-dashed line refers to the top-hat shape. Green dot-dashed, orange dashed, and red dotted lines corresponds to different values of the geometric index parameter $\alpha$ in equation (\ref{eq:mod_bump_function}): $\alpha=0.8, 0.4, 0.25$, respectively, where larger $\alpha$~values approach the top-hat shape. In the right panel of \autoref{fig:timeavg1D_compare}, we present the integrated results using these different shape functions, and compare them to the time-averaged simulated radial density profile at $\theta=0$ (black solid line).  The more top-hat shapes at high values of $\alpha$ have more laterally distributed density, which tends to result in slower transverse radiative diffusion.  Thermal equilibrium therefore forces them to have lower densities in the column compared to the shapes with small values of $\alpha$.  A value of $\alpha=0.40$ (orange dashed line) produces a mound shape that closely resembles that of the time-averaged simulated density profile, and the 1D radial profile is reasonably consistent with that of the simulation along the polar axis.  A top-hat geometry always overestimates the shock height because it underestimates the cooling efficiency.

\begin{figure*}
    \centering
	\includegraphics[width=\textwidth]{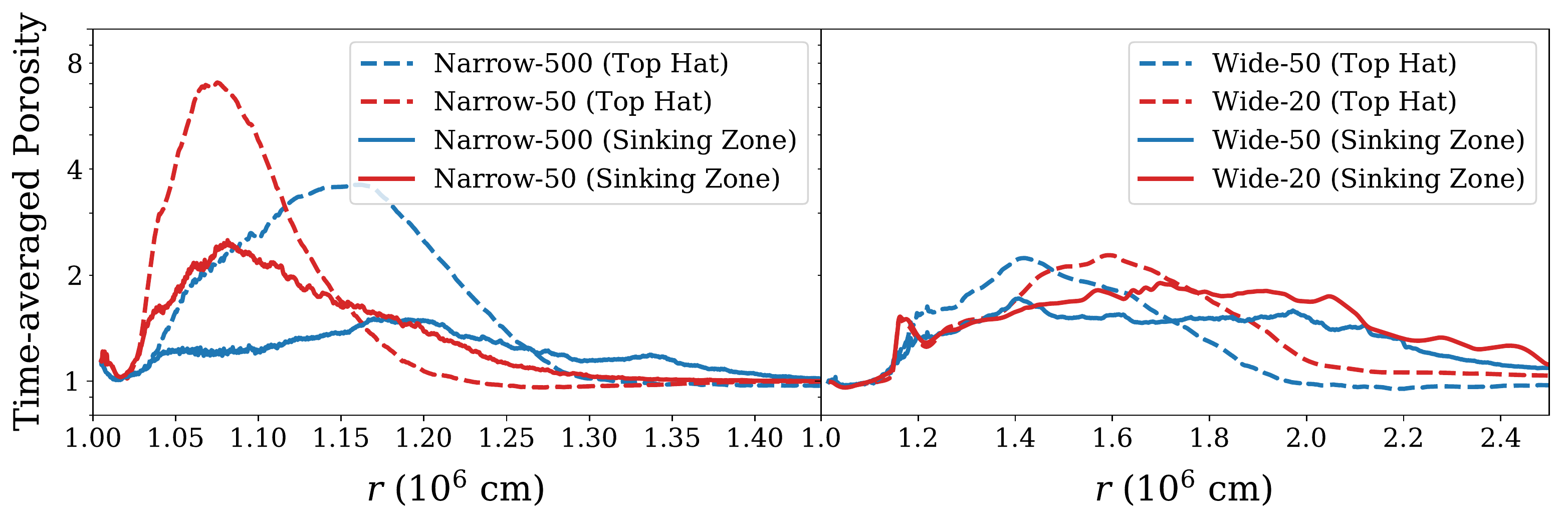}
    \caption{Time-averaged porosity as a function of radial position measured in our two narrow column simulations (left) and our two wide column simulations (right).  Solid curves show averages of the porosity below the instantaneous position of the shock, i.e. the transverse average at a particular radius and time extends from the polar axis to the portion of the shock which has that radius, or to the edge of the column.  Dashed curves show the porosity calculated from transverse averages from the polar axis to the edge of the outermost confining field line.  This can include portions of the free-fall region, but is the appropriate measure to use for 1D models with top hat column shapes.  The porosity is greater than unity throughout the sinking zone in all cases, indicating super-Eddington radial fluxes. Note that for Wide-20, we smooth the curve to remove the artificial discontinuities arising from finite time sampling over a relatively large amplitude oscillation.}
    \label{fig:porosity}
\end{figure*}

Note that the local Eddington ratio in the sinking zone is nearly zero in the time average (see the last panel of \autoref{fig:timeavg2D}), while the instantaneous radiation pressure support can exceed the local Eddington limit in certain locations (see e.g. the last panels of \autoref{fig:snapshot1_fiducial} or \autoref{fig:snapshot2_fiducial}).  This super-Eddington comoving radial flux is enabled by the density inhomogeneities within the accretion column, where the flux is larger in the lower density finger-shaped channels \citep{Begelman2001, Shaviv1998}.  To properly measure this effect, we define the porosity as: 
\begin{equation}
    \mathcal{P}(r,t) \equiv \frac{\left<\rho\kappa\right> \left<|\bar{F}_{r,r}|\right>}{\left<\rho\kappa|\bar{F}_{r,r}|\right>}
    \quad, 
\end{equation}
where $\bar{F}_{r,r}$ is the radial comoving radiation flux, and $\kappa$ is the flux mean opacity, which in our simulations is dominated by Thomson opacity.  The angle bracket represents a spatial average in the polar direction, which we do in two ways:  from the polar axis to the accretion shock or last confining field line, whichever comes first (i.e. just from the actual sinking zone), or from the polar axis to the last confining field line, even if this includes the free-fall region.  The latter would be appropriate for 1D models that assume a top hat geometry rather than the actual mound shape of the time-averaged column structure.  In \autoref{fig:porosity}, we display the time-averaged porosities computed in both ways as a function of radial position, for all four of our high-resolution simulations.  We generally find the porosity to be greater than unity in the sinking zone, consistent with super-Eddington radial fluxes.  Note, however, for the fiducial simulation, the porosity as computed just from the sinking zone is near unity at almost all radii, which explains why a simple hydrostatic model with the correct time-averaged mound shape produces a radial density profile that agrees with the simulation data (the $\alpha=0.4$ case in \autoref{fig:timeavg1D_compare}).  These porosity profiles could in principle be used to improve 1D models of accretion columns.

\subsection{Energetics of Accretion Columns}
\label{sec:energetics_of_accretion_columns}

\begin{figure*}
    \centering
	\includegraphics[width=0.9\textwidth]{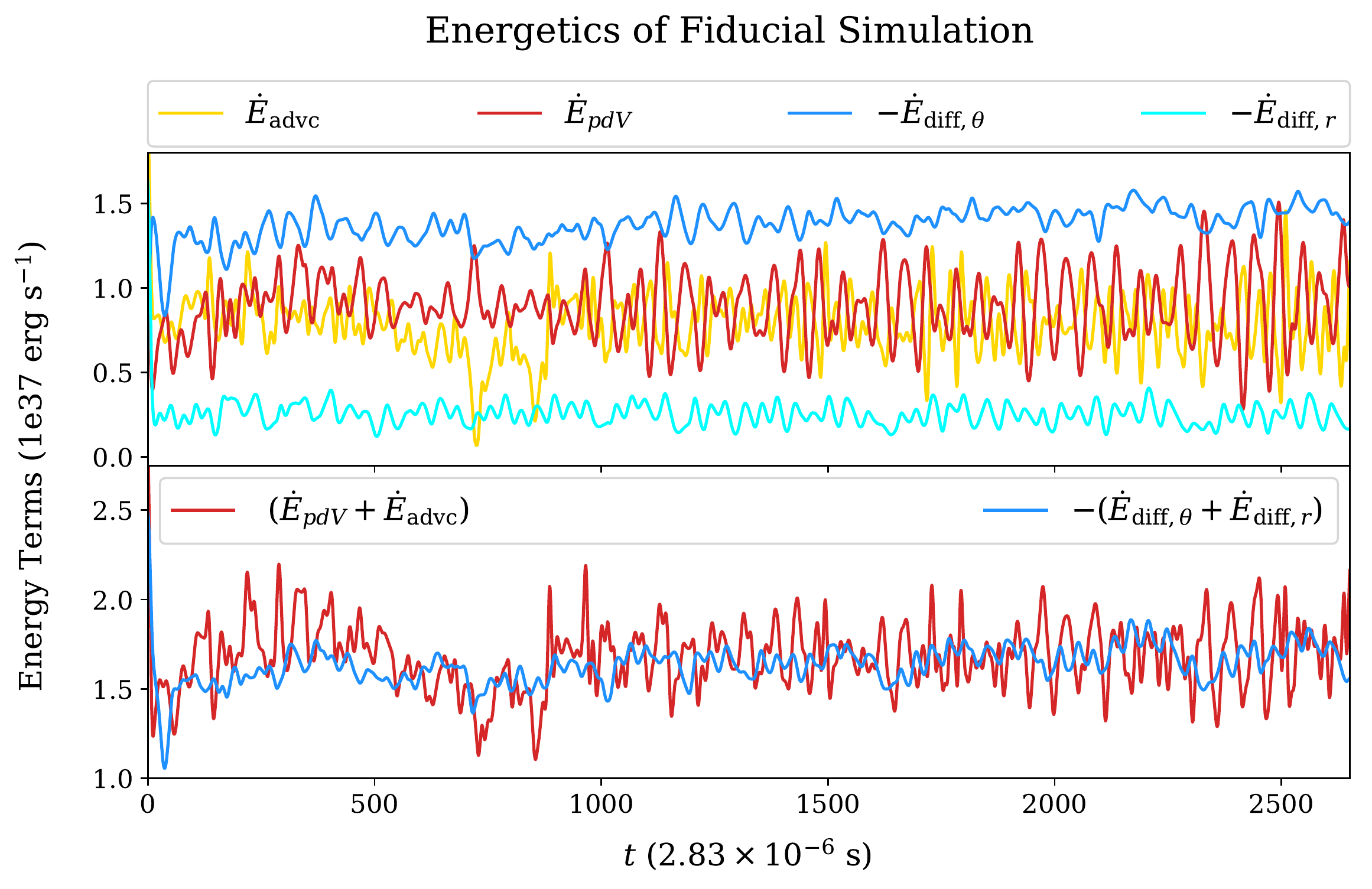}
    \caption{Energetics of the sinking zone in the fiducial simulation.  In the upper panel, the major cooling mechanism of the accretion column is the sideways radiation emission (blue curve), which completely dominates the emission in the radial direction (cyan curve).  Internal energy input comes from both advection (gold curve) and \textit{pdV} work (red curve).  The lower panel shows that all the heating and cooling terms achieve an approximate thermal equilibrium.  Note that we have smoothed the curves by box car time-averaging the data over 4$t_{\mathrm{sim}}$ ($\sim 10^{-5}$~s) in order to remove variability on time scales much shorter than the oscillation time scale of the column.
    }
    \label{fig:energetics_fiducial}
\end{figure*}

The structure of the accretion column is very sensitive to the balance between its heating and cooling processes, and we analyze these in detail here, focusing on the sinking zone.  Recall that the gas in the sinking zone moves much more slowly than the speed of light, so we can examine the rate of change of energy terms in the Newtonian regime.  These are, respectively, advection, \textit{pdV} work, transverse radiation diffusion, and radial radiation diffusion, and are defined as follows: 
\begin{subequations}
\begin{align}
    \dot{E}_{\mathrm{advc}} &= -\oint \left(\frac{P_g}{\gamma-1} + \bar{E}_r \right) v_r dA_r
    \quad, 
    \\
    \dot{E}_{\textit{pdV}} &= -\int \left(P_g+\bar{P}_r\right)\frac{\partial}{\partial r}\left(r^2 v_r\right) dV
    \quad, 
    \\
    \dot{E}_{\mathrm{diff}, \theta} &= -\oint \bar{F}_{r,\theta} dA_{\theta}
    \quad, 
    \\
    \dot{E}_{\mathrm{diff}, r} &= -\oint \bar{F}_{r,r} dA_r
    \quad.
\end{align}
\end{subequations}
Here $\bar{P}_r$ is the comoving radiation pressure and $\bar{F}_{r,\theta}$ refers to the transverse comoving radiation flux.  $dA_r$ and $dA_{\theta}$ refer to the differential area elements in the radial and polar directions, respectively.  The quantity $dV$ is the volume element in spherical polar coordinates.  More detail on how we compute these quantities can be found in \hyperref[sec:appendix_2]{Appendix B}.

In \autoref{fig:energetics_fiducial}, we analyze the energetics inside the sinking zone of the fiducial simulation as a function of time.  As shown in the upper panel of \autoref{fig:energetics_fiducial}, both the advection process and the \textit{pdV} work contribute to increasing the internal energy of the system.  Sideways diffusion of radiation is the dominant cooling process, and the emergent radiation is therefore in the fan-beam pattern.  As shown in the lower panel of \autoref{fig:energetics_fiducial}, the overall heating and cooling roughly reaches thermal equilibrium. 

In our Cartesian simulations of \citet{Paper2}, the accretion rates were moderate and the shock heights were much lower than here.  We found in those cases that the advection process always dominated over the \textit{pdV} work in increasing the internal energy. Our simulations in this paper develop taller column structures in spherical polar geometry at higher accretion rates.  The convergence of the magnetic confinement over a larger radial range results in a \textit{pdV} work that is now comparable and sometimes larger than the advection, which is more consistent with the physical picture described in the 1D stationary model by \citet{Basko1976}.  The ratio of time-averaged \textit{pdV} work to advection $\left<\dot{E}_{\textit{pdV}}\right>_t/\left<\dot{E}_{\mathrm{advc}}\right>_t$ is 1.17 in the fiducial simulation.  Version 1 has the same column width, but ten times less accretion rate and therefore a shorter column, so this ratio is reduced to 0.64.  For the wider column simulation Version~2, which has the same area-weighted accretion rate, this ratio increases to 3.84.  Reduction of the accretion rate of this wide column (Version~3) again decreases the ratio, in this case to 1.66. 

\begin{figure*}
    \centering
	\includegraphics[width=\textwidth]{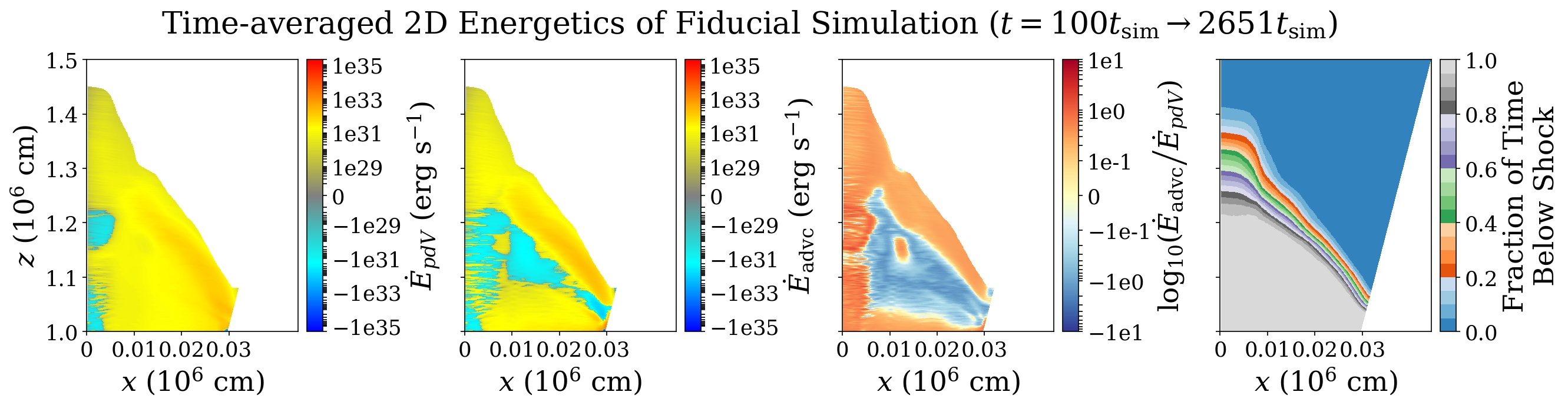}
    \caption{2D distribution of the rate of change of internal energy due to radial advection (left) and \textit{pdV} work (middle left), and their ratio (middle right).   The right-most panel shows the fraction of time a given point likes below the radially oscillating shock front.
    }
    \label{fig:advection_zone_fiducial}
\end{figure*}

The competition between radiative advection and \textit{pdV} work in supplying internal energy to balance cooling is important because it relates to the mechanism that drives the overall radial oscillations of the column structure.  In the short column, Cartesian simulations we presented in \citet{Paper2}, \textit{pdV} work was negligible and radiation advection brought energy inward from the shock front to balance cooling.  However, this balance was not achievable in a time-steady configuration, resulting in oscillations between phases where the column was tall and cooling exceeded heating, to phases where the column was short and heating exceeded cooling.  The presence of significant \textit{pdV} work in the simulations of this paper can provide a more local mechanism to balance cooling, just as in the original 1D stationary model of \cite{Basko1976}. \autoref{fig:advection_zone_fiducial} shows the time-averaged 2D structure of advection and \textit{pdV} work, as well as their ratio.  In the time average, the \textit{pdV} work dominates the radiation advection except at the top, on the polar axis, and near the neutron star surface.  

The advection dominated region near the shock originates from the shock oscillation, which is proved generating most of the light variation in \autoref{sec:luminosity_variation}.  The right-most 
panel of \autoref{fig:advection_zone_fiducial} shows the fraction of time during the shock oscillation that a particular point in space is below the accretion shock (i.e. 1 implies that the point is always below the shock front and 0 implies that the region is always radially outside the shock front).  It is particularly noteworthy that radiation advection dominates the rate of increase of internal energy precisely in the region where the shock is oscillating.  This strongly suggests that the oscillation that is responsible for the variations in the light curve in \autoref{sec:luminosity_variation} originates from exactly the same mechanism that we identified in \citet{Paper2}.  

In the bottom region, the \textit{pdV} work does not produce sufficient heat to support the base of the column and the advection must then be driven to transport the extra heat to the bottom, which is also indicated by the downward time-averaged lab-frame radiation flux (see the fourth panel of \autoref{fig:timeavg2D}).

\begin{figure*}
    \centering
	\includegraphics[width=0.85\textwidth]{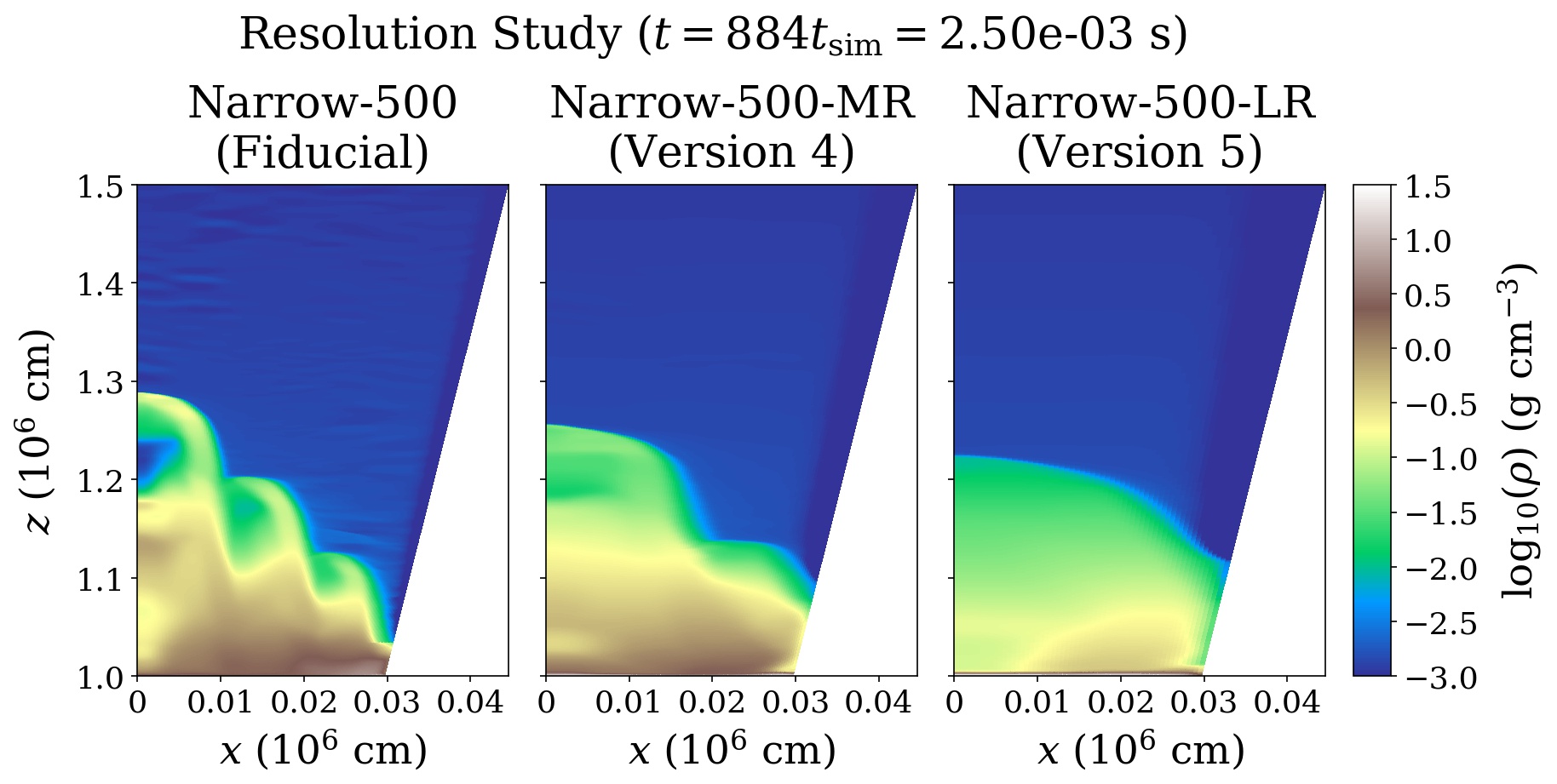}
    \caption{Snapshots of the density structure at $t=884t_{\mathrm{sim}}$ in our narrow column simulations at different resolutions from high (left) to low (right). }
    \label{fig:resolution_density}
\end{figure*}

Note from the first and second panels of \autoref{fig:advection_zone_fiducial} that both the \textit{pdV} work and the radiation advection along the polar axis below the shock oscillation region are cooling, not heating, in the time-average.  This is again related to the fact that the time-averaged, mass-weighted velocity near the lower polar region is outward, not inward (see second panel of \autoref{fig:timeavg2D}).  This also exacerbates the inability of the column to achieve a balance between heating and cooling, and may be a contributing factor to why the central finger structure along the polar axis has the largest oscillation amplitude. 

\subsection{Entropy Wave and Resolution Study}
\label{sec:entropy_wave_and_resolution_study}

As we noted above in \autoref{sec:build_up_of_accretion_column}, the wave pattern that propagates inward toward the polar axis strongly resembles a similar wave pattern that developed in our Cartesian simulations of \citet{Paper2}.  Because those simulations started with an initial condition consisting of an accretion column, we were able to show that the wave pattern resulted from the growth of the slow diffusion photon bubble instability (an unstable entropy mode).  Here, since the accretion column is built up dynamically from a free-fall flow in our new simulations, there is no steady background state in which to identify the growth of the unstable entropy mode. 

\lz{However, deep in the column, the wave amplitude is relatively small and arguably in the linear regime, and in the fiducial simulation it has short enough wavelengths to be in the WKB regime.  We can therefore measure its
phase velocity} in the simulation $v_{\mathrm{p, sim}}$, \lz{and compare it to} the expected value $v_{\mathrm{p, exp}}$ from the linear dispersion relation for the slow diffusion photon bubble instability (\citealt{Arons1992}, equation A22 in \citealt{Paper1}).  We measure the phase velocity in multiple epochs of the fiducial simulation by tracking a selected wave front propagating inward from the side toward the polar axis.  For the expected phase velocity, we solve the dispersion relation by first adopting a background state corresponding to the time average of the epochs over which we measure the simulated phase velocity. We did this by selecting the mode that had the maximum linear growth rate at the measured wavelength.  More detail about the measurement and calculation can be found in \hyperref[sec:appendix_3]{Appendix C}.  The results are summarized in the \autoref{tab:phase_velocity_entropy_wave}, where the simulation measurements and the theoretical expectations are very consistent.  This result is robust because there is no other characteristic speed near this regime. The sound speed is $\sim0.1c$ and the radiation diffusion speed is $\sim10^{-4}c$.  Therefore, the wave patterns that we observe in the simulation are indeed the entropy mode. 

\lz{Note that the local entropy wave dispersion relation that we use to calculate the phase velocity assumes a background with no transverse spatial variation, and therefore does not distinguish inward from outward transverse propagation.  We only observe inward transverse propagation from the sides to the center, and this is due to the fact that the sinking zone is mound-shaped, i.e. it is shorter on the sides than at the center.  Consider an outward radial motion of the side of the column.  This provides more shielding of the material immediately on the inside, reducing its sideways cooling, and causing it to expand radially in response as it overheats.  This in turn shields material further inward, causing it to expand radially.  Similarly, an inward radial oscillation at the sides provides less shielding of material on the inside, enhancing the sideways cooling and causing that material to contract radially.  All of this leads to inward propagation of the entropy wave.  Outward propagation does not occur because a radial expansion or contraction of the taller inner material has negligible effect on the shielding of the shorter material further out.}

\begin{table}
	\centering
	\begin{tabular}{cccccc}
		\hline
		$t/t_{\mathrm{sim}}$& 400 & 890 & 1500 & 2160 & 2510 \\
		& \textrightarrow520 & \textrightarrow1010 & \textrightarrow1620 & \textrightarrow2270 & \textrightarrow2630 \\
		\hline
		$v_{\mathrm{p, sim}}/c$ & 1.88e-3 & 2.14e-3 & 2.21e-3 & 2.30e-3 & 2.22e-3 \\
		$v_{\mathrm{p, exp}}/c$ & 1.91e-3 & 1.98e-3 & 2.07e-3 & 2.03e-3 & 2.14e-3 \\
		\hline
	\end{tabular}
	\caption{Comparison between the measured phase velocity and the theoretically expected phase velocity of the entropy wave in five different epochs of the fiducial simulation.  The simulation measurements are highly consistent with the expectation from the linear theory. }
	\label{tab:phase_velocity_entropy_wave}
\end{table}

Another characteristic of the entropy wave is its resolution dependence, as a higher resolution can result in a faster growing mode at the shorter resolved wavelength until the radiation viscous length scale \citep{Paper1,Paper2} is reached.  We have performed a resolution study on the fiducial simulation by running two additional simulations with decreased resolution by factors of 2 (Version~4) and 4 (Version~5).  The results resemble what we found in a similar resolution study in \citet{Paper2}.  As shown in \autoref{fig:resolution_density}, the number of fingers decreases by the same factor as the decrease in resolution, and the shape of the column becomes less peaked toward the polar axis.  Despite these changes with resolution, the global oscillation frequency remains roughly the same, only slightly increasing with decreasing resolution (\autoref{tab:qpo_characteristic}).  This robustness of the frequency with resolution is similar to what we found in the Cartesian simulations of \citet{Paper2}.  What increase in frequency is present appears to be \lz{due to} a smaller \lz{oscillating region} after losing the finger-shaped structure, \lz{and the fact that the presence of fingers decreases the overall cooling efficiency of the column}. In \citet{Paper2}, the dependence of the cooling on the horizontal shape of the \lz{Cartesian} column is less sensitive than in an axisymmetric column.  Having more resolved fingers in the Cartesian geometry therefore does not affect the cooling as much as it does here in the axisymmetric geometry.  Even with this increased sensitivity, the oscillation frequency depends only weakly on resolution.

\section{Discussion}
\label{sec:discussion}

\subsection{Numerical Caveats}
\label{sec:numerical_caveats}

There exist some numerical caveats in our simulations. We discuss them in detail here.  

\subsubsection{Geometric Dilution and Resolution}
The spherical polar coordinate grid that we use here has decreasing resolution with radius, and this might affect the growth and propagation of the entropy waves at different altitudes.  As shown in \citet{Paper1}, the slow-diffusion photon bubble instability that arises from the entropy wave grows faster at shorter wavelengths, until the radiation viscous length scale is resolved.  The viscous length scale in the sinking zone of the fiducial simulation ranges from $\sim 400$~cm at the stellar surface to $\sim2000$~cm at $r=1.2R_{\star}$, and requires at least $\sim50$ grid cells to resolve.  So in fact our grid is closest to the viscous length scale at larger radii. 

\subsubsection{Bottom Boundary Condition}
Recall that we adopt an effective bottom boundary that mocks up the neutron star surface layer.  In our simulations, the neutron star surface is heated via Compton scattering by energetic photons generated inside the sinking zone. In reality, such heat should flow into the neutron star and establish a thermal equilibrium with the accretion column.  However, since we adopt a reflective
boundary condition below the neutron star surface layer, the heat accumulates
and can eventually lead to an increased temperature of the bottom layer, which might overheat the simulated accretion column.  Thus, we set up a very thick effective bottom boundary in our simulations so that the heat capacity is large enough that the temperature increases only very slowly after the sinking zone is built up.  Note that we assume a nondegenerate ideal gas equation of state (EOS) for the effective bottom boundary, which overestimates the heat capacity and, as we already stated, underestimates the heat conduction into the neutron star.
This can obviously be further improved by properly treating the thermal properties of the neutron star surface with the correct EOS (e.g. \citealt{Negele1973}).  

\subsubsection{Dependence on the Magnetic Field}

As in \citet{Paper2}, we adopt a surface magnetic field strength of $10^{11}$~G in our simulations, which is lower than the typical high-mass X-ray binary pulsars ($\gtrsim10^{12}$~G).  However, as long as the magnetic field is strong enough to confine the transverse motion of the gas, it does not alter the dynamics of the accretion column.  However, this would not be true if we accounted for the dependence of opacity on magnetic field strength, rather than the simple Thomson opacity that we have assumed here.  The photon energy and polarization-averaged flux-mean opacity can be significantly reduced below Thomson for temperatures below the cyclotron energy (e.g. \citealt{Arons1992}).  Incorporation of magnetic opacity effects can therefore reduce the height of the column, and may significantly effect its dynamics.  These effects will be explored in our next paper.  

\subsection{Relationship of Oscillations to the Entropy Waves}
\label{sec:unstable_entropy_mode}

\lz{In \autoref{sec:entropy_wave_and_resolution_study} we demonstrated that the entropy waves that are present in the fiducial simulation have phase speeds that are consistent with the linear, slow diffusion photon bubble instability dispersion relation.  However, this does not explain the frequencies of the entropy waves that are actually present.  In particular,
the frequency of arrival of successive wave fronts of the entropy wave can at times be remarkably close to the oscillation frequency of the central finger-shaped structure along the polar axis.  Nevertheless, as we showed in our
Cartesian simulations of \citet{Paper2}, that oscillation frequency is determined simply by the overall net cooling time of the column.}

\lz{In fact,} recall from our resolution study in \autoref{sec:entropy_wave_and_resolution_study} \lz{that} the dominant frequency of the light curve is only weakly affected by the entropy waves, the small dependence being due to the effect of these waves on the overall shape of the column, which slightly alters its cooling rate.  \lz{At the same time, the different oscillating fingers themselves have higher frequencies as one moves outward from the center, because they have individual cooling times that are shorter.  A close examination of the arrival frequencies of successive \lz{entropy} wave fronts toward the center in the fiducial simulation show that they} arrive with frequencies ranging from 5.76 to 8.46~kHz, close to the QPO frequency:  the peak at $f_{\mathrm{peak}}=5.86$~kHz and the edge of the extended plateau $f_{\mathrm{break}}=9.66$~kHz.  We therefore suggest that the entropy waves \lz{that are present in the saturated state are in frequency} resonance with the \lz{radial oscillation frequencies of the individual fingers}.  This is also consistent with the behavior of the wider column simulations, which have broader QPO plateaus, and appear to have multiple entropy wave modes that are excited. 

\subsection{Comparison with Previous Works}
\label{sec:comparison_with_previous_works}

Our accretion column simulations develop finger-shaped structures (which might be called 'photon bubbles') introduced by the unstable entropy waves, which appear to resemble what \citet{Klein1989} and \citet{Klein1996}, found in their 2D simulations.  In particular, they share similar high-frequency oscillatory behavior.  \citet{Klein1989} and \citet{Klein1996} attributed these oscillations to photon bubbles, but our resolution studies (section~3.4 in \citealt{Paper2} and \autoref{sec:entropy_wave_and_resolution_study}) suggest that the oscillatory behaviors arise instead from the inability of the column to maintain a stationary balance between accretion power heating and radiative cooling (see section~3.3 in \citealt{Paper2} and \autoref{sec:energetics_of_accretion_columns}).  Note that \citet{Klein1989} observed mergers of their photon bubbles toward larger spatial structures.  We see similar behavior in our wide column simulations, in that the accretion shock forms wide localized mound shapes, inside of which the shorter wavelength entropy waves still propagate.  We remind the reader that the spatial scales of the entropy wave are in fact resolution-dependent.  However, this has only a small effect on the QPO frequency.  

\citet{Kawashima2020} have also done simulations of neutron star accretion columns in monopole geometry.  Their initial and boundary conditions differ from ours, as they start with an extended, uniform mass distribution out of hydrostatic equilibrium that then free-falls onto the neutron star surface.  In their simulations, the flow breaks up into narrow, dense channels that accrete and surround broad, underdense regions that undergo radiation pressure-driven outflow.  They do not find evidence of a quasi-stationary, oscillating structure such as we find.  This could be due to the fact that their accretion luminosity exceeds the global Eddington luminosity of the entire neutron star by large factors, much larger than the sub-Eddington luminosities of our simulations here. 

\citet{Abolmasov2022} recently reported time-dependent, one-dimensional simulations of accretion columns in a dipolar magnetic geometry.  Intriguingly, they found that their accretion shocks at the top of the optically thick columns exhibited radial high frequency oscillations, though in contrast to what we find here, their shock oscillations were damped with time.  Their oscillation period is close to the radial sound propagation time through the column, whereas our oscillations are nearly hydrostatic.  Hence it appears that they have found a damped acoustic oscillation.  The reason why they did not find the longer time scale thermal oscillation that we have found here may be due to their 1D approximation.  Our column has a changing shape which is overheated at its most compressed state, then overcooled at its most expanded state because it has a larger sideways cooling area.  This effect of changing shape would not be accounted for in their 1D model. 

In our previous Cartesian simulations \citep{Paper2}, we purposely limited the accretion rate so that the column height never became a significant fraction of the neutron star radius.  Here the split monopole magnetic geometry has allowed us to break this constraint and move to higher luminosities with some realism, although the field does not diverge as quickly as, say, a more realistic dipole or higher order multipole field.  There are two main physical differences between our previous short-column, Cartesian simulations and the current, taller and geometrically diluted simulations.  First, we find no prominent pre-shock behavior in the fiducial simulation here, even though pre-shocks were prominent in the Cartesian simulations.  This may be because the density in the free-fall region decreases more rapidly with radius in the diverging magnetic field, so that there is less ram pressure in the free-fall region.  On the other hand, simulation Wide-20 (Version 3) does show occasional shock structures near the polar axis in the free-fall region.

Second, \textit{pdV} work was a negligible contribution to increasing internal energy in the sinking zone of the short Cartesian simulations. Instead, radiation advection was the dominant source of increasing internal energy, but this transport mechanism downward from the accretion shock was unable to balance radiative cooling in a stationary fashion, and this is what drove the oscillations.  In the taller simulations here, \textit{pdV} work plays an increasingly important role, both with increasing accretion rate and column height.  This is because of the sideways compression that is present in the split monopole geometry, compared to Cartesian simulations at comparable local Eddington ratios.  The latter fact suggests that it would be even more important in dipole and quadrupole magnetic field geometries.  It is possible that tall accretion columns in more diverging magnetic field geometries might achieve an equilibrium between \textit{pdV} work and radiative cooling, akin to the physical picture in the 1D model of \citet{Basko1976}.  However, that column would still be unstable to the slow-diffusion photon bubble instability.  Moreover, accounting for magnetic opacities will result in shorter columns, and we will explore this effect in our next paper.  

Finally, we emphasize again that that the 2D shape of the column determines the cooling efficiency, and radial profiles of various quantities within the sinking zone depend sensitively on this.  The column also has significant porosity.  We have demonstrated in \autoref{sec:time_averaged_profiles} how these effects can in principle be accounted for in 1D models. 

\subsection{Observational Significance and Future Work}
\label{sec:observational_significance_and_future_work}

Compared to the Cartesian simulations of \citet{Paper2}, we are now in the accretion regime of many X-ray pulsars and even the low luminosity end of ULXs.  These new simulations are therefore more useful for comparing with observations.  In agreement with the pioneering work of \citep{Klein1996}, our new simulations strongly suggest that high luminosity X-ray pulsars should exhibit high frequency oscillations (2-10 kHz, depending on parameters).  Detection of such high frequencies will likely be challenging with existing X-ray timing facilities, and indeed has a somewhat checkered history \citep{Jernigan2000,Revnivtsev2015}, but we still feel that attempts are worthwhile given how robust is this physics.

An additional observable effect might be the effects of the changed time-averaged structure of the column, compared to 1D models, on predicted spectra and polarization.  Our simulations can be post-processed to make such predictions.  \lz{Note that the bulk velocities in the column are less than the electron thermal speed by factors of two to one hundred, with a median value of around ten.  We therefore expect thermal Comptonization to dominate over bulk Comptonization, and this will enhance the thermalization of the emergent radiation with the gas temperature.}

Finally, there is currently a debate in the ULX community concerning how their apparent super-Eddington luminosities can be achieved. One way is to have a strong magnetic field in the neutron star leading to a significant magnetic reduction of opacity.  Such low magnetic opacity can largely reduce the gas-radiation interaction and permit a much higher accretion rate within the accretion column \citep{Mushtukov2015}.  Another way is to have strong geometric beaming from the surrounding optically thick accretion flow \citep{Abarca2021}, though this might result in the smearing out of the observed neutron star spin pulsations \citep{Mushtukov2021}.  It will be important to test this claim with future numerical simulations that include the more global structure of the accretion flow.

\section{Conclusions}
\label{sec:conclusions}

We summarize our conclusions as follows:
\begin{enumerate}
    \item The light curves in all of our simulations display high-frequency ($2-10$~kHz) QPOs.  These QPOs are dominated in the power spectrum by a peak frequency originating from the oscillating polar axis region of the accretion column, but include an extended plateau toward higher frequencies due to the higher frequency shock oscillations further out from the polar axis.  The peak frequency and fractional amplitude depend on the global parameters of the accretion column, being lower for higher accretion rates or wider columns. 
    
    \item We have confirmed that when the accretion rate is supercritical, a radiation-pressure dominated sinking zone is formed as predicted by \citet{Basko1976}.  In our simulations, the sinking zone appears to be mound-shaped, which can significantly affect the cooling efficiency and thus modify the radial structure of the accretion column.  We demonstrate the consistency between the time-averaged profile of the fiducial simulation and the 1D stationary model provided it accounts for the correct shape function.  Accounting for porosity (which is not that different from unity in the fiducial simulation) should also improve 1D models. One final effect is that the oscillations smooth out the shock structure in the time-average, an effect which cannot be accounted for in 1D models that assume shock jump conditions. 
    
    \item In our simulations, the accretion column is mainly cooled by sideways emission of radiation. This cooling is balanced by increases of internal energy due to \textit{pdV} work and radial radiation advection.  The \textit{pdV} work arises in part from the sideways compression of the converging magnetic field toward the neutron star surface.  Radiation advection originates mostly from the radial oscillations within the column.  When the accretion column is wider or has higher accretion rate, \textit{pdV} work becomes more important compared to advection.  We expect that \textit{pdV} work will be even more important for dipolar or quadrupolar field geometries.  

    \item We observe the entropy wave that is associated with the slow-diffusion photon bubble instability in our simulations.  While the properties of this wave depends on numerical resolution, we confirm that the dominant oscillation frequency depends only weakly on these properties.  This is consistent with what we found in \citet{Paper2}.  What effects the entropy wave has on the oscillation are due to the altered mound shape of the column and resulting change in cooling efficiency. 
    
\end{enumerate}

\section*{Acknowledgements}

We thank Mitch Begelman, Matt Middleton, Xin Sheng, and Jim Stone for useful conversations, \lz{and the referee for very helpful remarks that significantly improved this paper}.  This work was supported in part by NASA Astrophysics Theory Program grant 80NSSC20K0525.  Resources supporting this work were provided by the NASA High-End Computing (HEC) Program through the NASA Advanced Supercomputing (NAS) Division at Ames Research Center.  We also used computational facilities purchased with funds from the National Science Foundation (CNS-1725797) and administered by the Center for Scientific Computing (CSC). The CSC is supported by the California NanoSystems Institute and the Materials Research Science and Engineering Center (MRSEC; NSF DMR 1720256) at UC Santa Barbara. The Center for Computational Astrophysics at the Flatiron Institute is supported by the Simons Foundation.

\section*{Data Availability}

All the simulation data reported here is available upon request to the authors.



\bibliographystyle{mnras}
\bibliography{references} 




\appendix

\section{Incorporating Distinct Mound Shapes in 1D Stationary Models}
\label{sec:appendix_1}

\begin{figure}
    \centering
	\includegraphics[width=0.8\columnwidth]{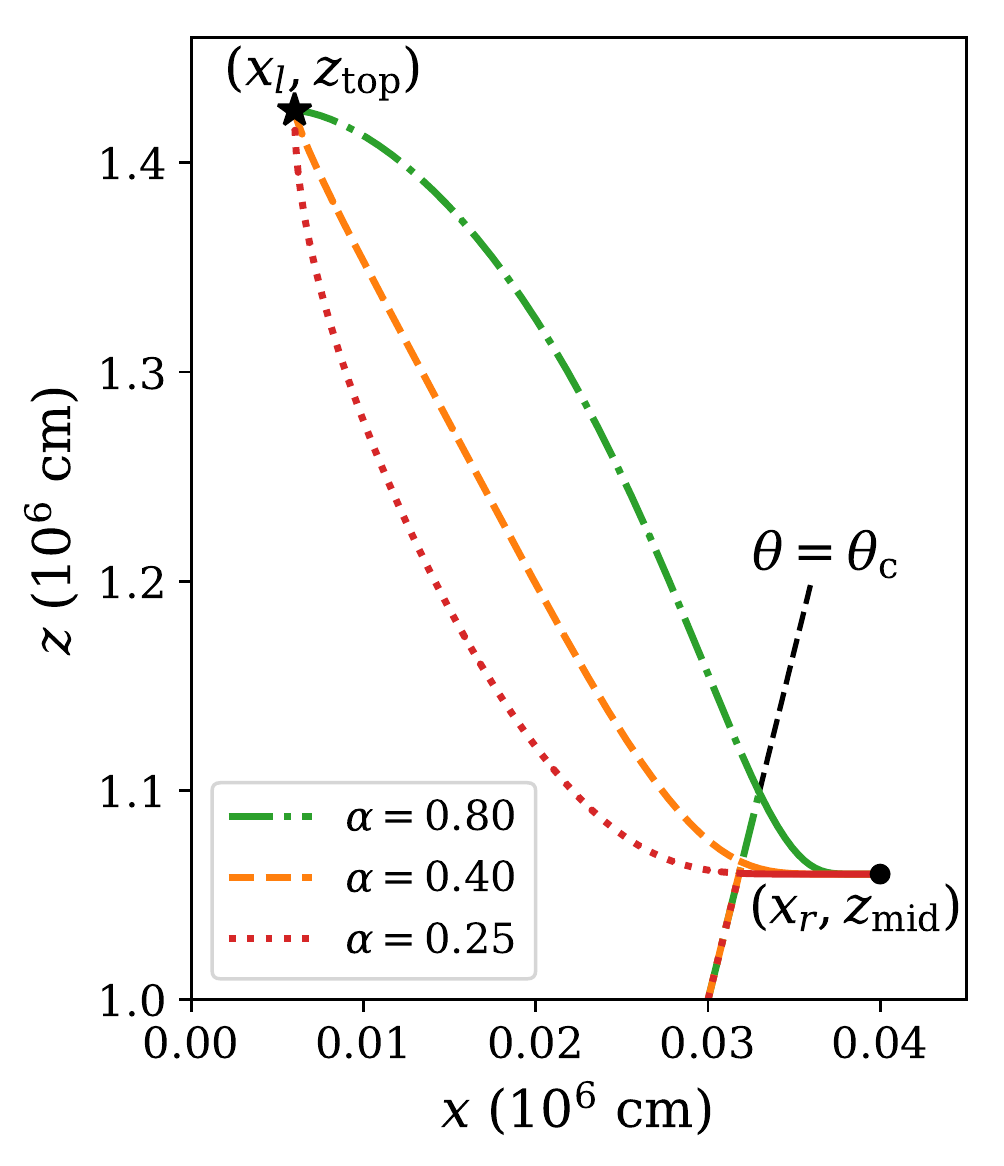}
    \caption{The shape function of equation \autoref{eq:mod_bump_function} with fixed parameters $x_l, x_r, z_{\mathrm{top}}$, and $z_{\mathrm{mid}}$, and varying geometric indices $\alpha$ as indicated. 
    }
    \label{fig:shape_function}
\end{figure}

Following section~2.3 in \citet{Paper2}, we adopt the same assumptions and approximations to formulate the 1D stationary model in spherical polar coordinates with a mound-shaped column structure of the sinking zone.  The equation set for the numerical integration is similar to what we derived in Cartesian coordinates (equations~11 in \citealt{Paper2}) but with extra geometric terms from the spherical polar geometry.  Additionally, we modify the one-zone approximation of the transverse radiation diffusion in the energy equation to account for the axisymmetry. The radial profiles of fluid quantities are then give by the following ordinary differential equations: 
\begin{subequations}
\begin{align}
    \frac{\partial T}{\partial r} &= -\frac{3\rho\kappa}{4a_r T^3} \frac{\bar{F}_{r,r}}{c}
    \quad, 
    \\
    \frac{\partial v}{\partial r} &= \dfrac{\left(g - \kappa\dfrac{\tilde{F}_{r,r}}{c} + R\dfrac{\partial T}{\partial r} - \dfrac{2RT}{r}\right)v}{RT-v^2}
    \quad, 
    \\
    \frac{\partial\rho}{\partial r} &= -\frac{\rho}{v}\frac{\partial v}{\partial r} - \frac{2\rho}{r}
    \quad, 
    \\
    \frac{\partial \bar{F}_{r,r}}{\partial r} &= 
    -\frac{2ca_r T^4}{3\rho\kappa x^2} - \frac{vRT}{\gamma-1}\frac{\partial\rho}{\partial r} - v\left(\frac{\rho R}{\gamma-1} + 4a_r T^3\right)\frac{\partial T}{\partial r} 
    \nonumber
    \\
    &\mkern17mu - \left(\frac{\gamma \rho RT}{\gamma-1} + \frac{4}{3}a_r T^4 \right)\left(\frac{\partial v}{\partial r} + \frac{2v}{r}\right) - \frac{2 \bar{F}_{r,r}}{r}
    \quad,  
\end{align}
\end{subequations}
where $a_r$ and $R$ refer to the radiation density constant and the ideal gas constant, respectively.  Note that we assume local thermal equilibrium here so that $T$ represents both the gas and radiation effective temperature.  The quantity $x$ refers to the transverse distance from the center to the side of the column, which can be implemented with a modified shape function to account for the 2D mound shape within the one-zone model: 
\begin{equation}
    x(z)=\begin{dcases}
        (x_r-x_l)\left[\left(\log\left(\dfrac{z-z_{\mathrm{mid}}}{z_{\mathrm{top}}-z_{\mathrm{mid}}}\right)-1\right)^{-1}+1\right]^{\frac{1}{2\alpha}} + x_l
        \quad,
        \\
        \mkern275mu z_{\mathrm{mid}} < z \le z_{\mathrm{top}}
        \\
        z\tan\theta_{\mathrm{c}} \quad, \quad R_{\star} \le z \le z_{\mathrm{mid}}
    \end{dcases}
    \label{eq:mod_bump_function}
\end{equation}
where $x=r\sin\theta$ is a function of $z=r\cos\theta$.  The geometry index $\alpha$ determines the shape of the mound:  large values of $\alpha$ result in a top-hat shape while small values of $\alpha$ produce a shape that is more concentrated toward the polar axis.  The other parameters are illustrated in \autoref{fig:shape_function}.  Moving outward from the polar axis, we take the height $z$ of the mound shape to be constant until the position ($x_l, z_{\mathrm{top}}$) is reached.  The height of the mound then declines according to our shape function, until it becomes flat again at position ($x_r, z_{\mathrm{mid}}$).  This latter position generally lies outside the column, as we cut off the function at the outermost radial field line at polar angle $\theta_{\mathrm{c}}$.  In \autoref{sec:time_averaged_profiles}, we adopt fixed values of $x_l=0.006R_{\star}$, $x_r=0.04R_{\star}$, $z_{\mathrm{mid}}=1.06R_{\star}$, and $z_{\mathrm{top}}=1.425R_{\star}$ for the shape function.  We only vary the parameter $\alpha$;  the values $\infty$, 0.8, 0.4 and 0.25 correspond to the cyan dot-dot-dashed, green dot-dashed, orange dashed, and red dotted curves depicted in \autoref{fig:timeavg1D_compare}, respectively. 

\section{Computing Rate of Change of Energy Terms}
\label{sec:appendix_2}

In neutron star accretion columns, the gas is well-confined by the magnetic field and the sinking zone is optically thick everywhere.  For the purposes of analyzing the energetics of the sinking zone, we can therefore make the simplifying assumptions of 1D gas motion along the radial (magnetic field line) direction, Eddington closure of the radiation moments, and the radiation diffusion approximation.  When the accretion column reaches a steady state, we can further ignore the time variation, and the Newtonian energy conservation equation becomes
\begin{subequations}
\begin{equation}
    \nabla_r\left(E_{\mathrm{tot}}v_r\right) + P_{\mathrm{tot}}\nabla_r v_r + \nabla_j\bar{F}^j = 0
    \quad.
    \label{eq:energy_balance_diff}
\end{equation}
Here $\nabla_r$ refers to the radial component of the divergence operator, $E_{\mathrm{tot}}=(\gamma-1)^{-1}P_g+\bar{E}_r$ is the total internal energy of the gas and radiation, and $P_{\mathrm{tot}}=P_g+\bar{P}_r$ is the total pressure in the fluid rest frame. By volume-integrating equation~\autoref{eq:energy_balance_diff} in the sinking zone, we obtain
\begin{equation}
    \dot{E}_{\mathrm{advc}} + \dot{E}_{\textit{pdV}} + \dot{E}_{\mathrm{diff}, \theta} + \dot{E}_{\mathrm{diff}, r} = 0
    \quad.
    \label{eq:energy_balance_intg}
\end{equation}
\end{subequations}
The four terms here refer to the rate of change of energy due to advection, \textit{pdV} work, transverse radiation diffusion, and radiation diffusion in the radial direction, respectively.  We compute these quantities as follows: 
\begin{subequations}
\begin{align}
    \dot{E}_{\mathrm{advc}} &= -\oint E_{\mathrm{tot}}v_r dA_r = -\sum_{\mathtt{i,j}} E_{\mathrm{tot}}^{(\mathtt{i,j})} v_r^{(\mathtt{i,j})} \Delta A_r^{(\mathtt{i,j})}
    \quad, 
    \\
    \dot{E}_{\textit{pdV}} &= -\int P_{\mathrm{tot}}\frac{\partial}{\partial r}\left(r^2 v_r\right) dV
    \nonumber
    \\
    &= -\sum_{\mathtt{i,j}} P_{\mathrm{tot}}^{(\mathtt{i,j})} \frac{r_{\mathtt{i}+1}^2 v_r^{(\mathtt{i}+1,\mathtt{j})} - r_{\mathtt{i}-1}^2 v_r^{(\mathtt{i}-1,\mathtt{j})}}{r_{\mathtt{i}+1} - r_{\mathtt{i}-1}} \Delta V^{(\mathtt{i,j})}
    \quad, 
    \\
    \dot{E}_{\mathrm{diff}, \theta} &= -\oint \bar{F}_{r,\theta} dA_{\theta} = -\sum_{\mathtt{i,j}} \bar{F}_{r,\theta}^{(\mathtt{i,j})} \Delta A_{\theta}^{(\mathtt{i,j})}
    \quad, 
    \\
    \dot{E}_{\mathrm{diff}, r} &= -\oint \bar{F}_{r,r} dA_r = -\sum_{\mathtt{i,j}} \bar{F}_{r,r}^{(\mathtt{i,j})} \Delta A_r^{(\mathtt{i,j})}
    \quad,
\end{align}
\end{subequations}
where the Latin indices $\mathtt{i}$ and $\mathtt{j}$ in typewriter font refer to the indices of the grid cell coordinates in $r$ and $\theta$, respectively.  Integration/summation is done with respect to the volume $V$ and area elements $A_r$ and $A_{\theta}$ that are orthogonal to the $\hat{r}$ and $\hat{\theta}$ directions, respectively.  The discrete forms of the grid cell interfaces and volume are
\begin{subequations}
\begin{align}
    & \Delta A_r^{(\mathtt{i,j})} = 2\pi r_{\mathtt{i}}^2 (\cos\theta_{\mathtt{j}-\frac{1}{2}} - \cos\theta_{\mathtt{j}+\frac{1}{2}})
    \quad, 
    \\
    & \Delta A_{\theta}^{(\mathtt{i,j})} = \pi\sin\theta_{\mathtt{j}}(r_{\mathtt{i}+\frac{1}{2}}^2-r_{\mathtt{i}-\frac{1}{2}}^2)
    \quad, 
    \\
    & \Delta V^{(\mathtt{i,j})} = \frac{2}{3}\pi(\cos\theta_{\mathtt{j}-\frac{1}{2}}-\cos\theta_{\mathtt{j}+\frac{1}{2}})(r_{\mathtt{i}+\frac{1}{2}}^3-r_{\mathtt{i}-\frac{1}{2}}^3)
    \quad, 
\end{align}
\end{subequations}
where the integer and half-integer indices refer to the cell center and the cell faces, respectively.

\section{Measurement and Analytical Solution of Phase Velocity of Entropy Wave}
\label{sec:appendix_3}

\begin{figure*}
    \centering
	\includegraphics[width=\textwidth]{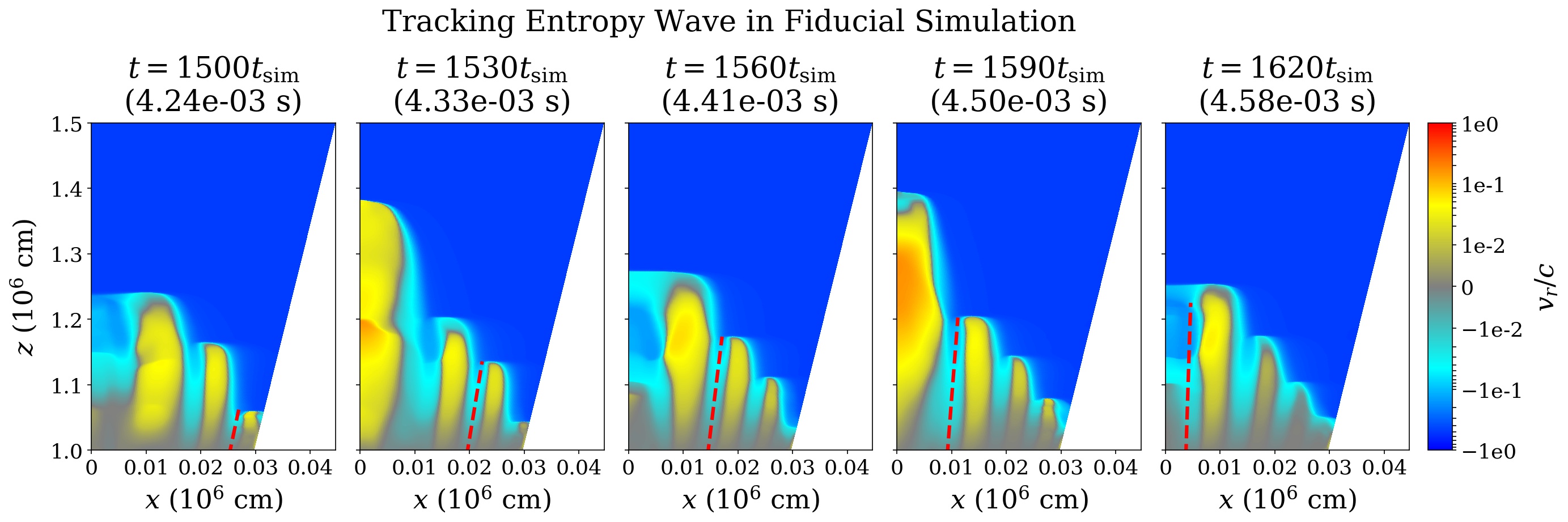}
    \caption{Snapshots of radial velocity in the fiducial simulation, illustrating how one can track the propagation of a particular minimum in the radial velocity of the wave pattern, indicated by the red dashed lines. 
    }
    \label{fig:wave_propagate_fiducial}
\end{figure*}

\begin{figure*}
    \centering
	\includegraphics[width=\textwidth]{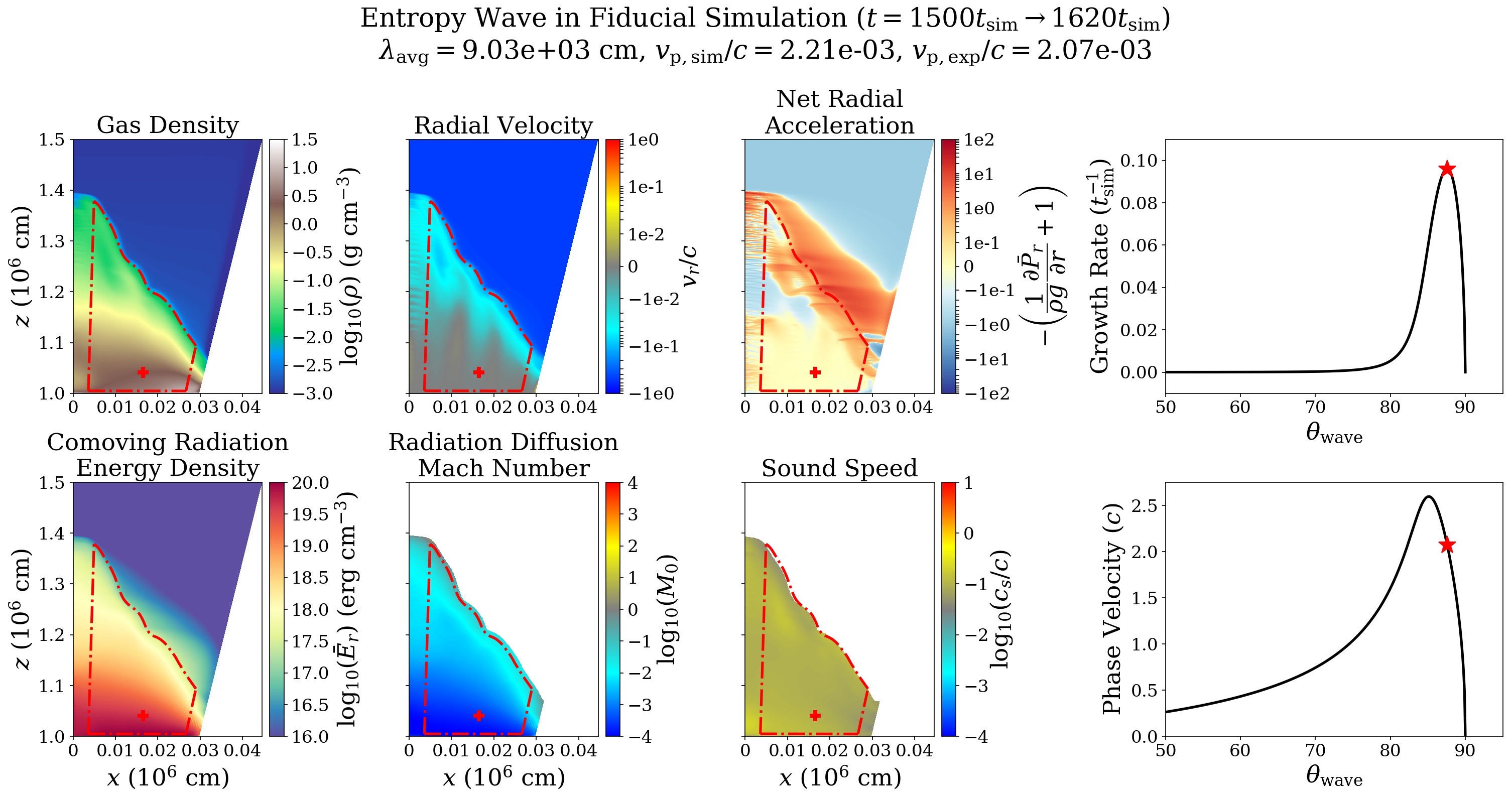}
    \caption{Related measurements to compute the expected phase velocity.  We adopt time-averaged profiles over the selected time range of entropy wave propagation as a background state for solving the linear dispersion relation.  The red dot-dashed lines enclose the region we use for our computation (i.e. the sinking zone that through which the entropy mode propagates during the selected range of time).  The red cross indicates the radiation energy-density weighted center of the enclosed region of the accretion column.   At the measured wavelength, we select the expected phase velocity for the maximum growing mode, which is indicated by the red stars in the right two panels, the upper showing growth rate and the lower showing phase velocity, as a function of wave propagation direction.}
    \label{fig:wave_measure_fiducial}
\end{figure*}

To measure the phase velocity of the entropy wave, we simply track a particular local minimum in the wave pattern of radial velocity moving in the 2D radial velocity profile, as illustrated in \autoref{fig:wave_propagate_fiducial}.  To calculate the expected phase velocity from linear theory, we first adopt time-averaged profiles of the sinking zone over the range of time that we use to track the wave propagation.  Then we compute the volume average of various fluid quantities over the region through which the simulated wave propagates and use them to solve the linear dispersion relation (equation~A22 in \citealt{Paper1}).  The angular frequency of the entropy mode depends on both the wave propagation direction and the wavelength.  For the wave propagation direction, we directly adopt the angle that maximizes the imaginary part of the frequency in order to select the dominant unstable mode.  For the wavelength, we time-average the wavelength measured from simulation snapshots over the time range that we use.  Note that the comparison of the phase velocity requires specification of the altitude, and we simply take the radiation energy density-weighted altitude.  In \autoref{fig:wave_measure_fiducial}, we present the time-averaged 2D profiles and the corresponding maximum growth rate and phase velocity.  The red dot-dashed lines indicate the sinking zone through which the entropy wave propagates, and the red cross inside this region represents the center of internal radiation energy.  In the last column of \autoref{fig:wave_measure_fiducial}, red stars refer to the maximum growing entropy mode and its phase velocity according to the measured averaged wavelength.


\bsp	
\label{lastpage}
\end{CJK*}
\end{document}